\pdfoutput=1

\documentclass[sigconf]{aamas}

\usepackage{balance} 

\usepackage[utf8]{inputenc} 
\usepackage[T1]{fontenc}    
\usepackage{hyperref}       
\usepackage{url}            
\usepackage{booktabs}       
\usepackage{amsmath}
\usepackage{amsfonts}       
\usepackage{nicefrac}       
\usepackage{microtype}
\usepackage{wrapfig}
\usepackage{graphicx}
\usepackage{subcaption}

\usepackage{xcolor}

\newcommand{\tsc}[1]{\textsuperscript{#1}}

\title{Modelling Cooperation in Network Games \\ with Spatio-Temporal Complexity}

\author{Michiel A. Bakker\tsc{1,2}*, Richard Everett\tsc{1}*, Laura Weidinger\tsc{1}, Iason Gabriel\tsc{1}, William S. Isaac\tsc{1}, Joel Z. Leibo\tsc{1}, \& Edward Hughes\tsc{1}}
\affiliation{
  \institution{1. DeepMind, 2. MIT}}
\email{{bakker@mit.edu,{reverett,lweidinger,iason,jzl,williamis,edwardhughes}@google.com}}
\thanks{*these authors contributed equally}

\begin{abstract}
The real world is awash with multi-agent problems that require collective action by self-interested agents, from the routing of packets across a computer network to the management of irrigation systems. Such systems have local incentives for individuals, whose behavior has an impact on the global outcome for the group. Given appropriate mechanisms describing agent interaction, groups may achieve socially beneficial outcomes, even in the face of short-term selfish incentives. In many cases, collective action problems possess an underlying graph structure, whose topology crucially determines the relationship between local decisions and emergent global effects. Such scenarios have received great attention through the lens of network games. However, this abstraction typically collapses important dimensions, such as geometry and time, relevant to the design of mechanisms promoting cooperation. In parallel work, multi-agent deep reinforcement learning has shown great promise in modelling the emergence of self-organized cooperation in complex gridworld domains. Here we apply this paradigm in graph-structured collective action problems. Using multi-agent deep reinforcement learning, we simulate an agent society for a variety of plausible mechanisms, finding clear transitions between different equilibria over time. We define analytic tools inspired by related literatures to measure the social outcomes, and use these to draw conclusions about the efficacy of different environmental interventions. Our methods have implications for mechanism design in both human and artificial agent systems.   
\end{abstract}

\newcommand{\BibTeX}{\rm B\kern-.05em{\sc i\kern-.025em b}\kern-.08em\TeX}

\begin{document}

\pagestyle{fancy}
\fancyhead{}

\maketitle

\section{Introduction}

Collective action problems are ubiquitous in the modern world and represent an important challenge for artificial intelligence. Routing packets across computer networks~\cite{DBLP:journals/corr/abs-1106-1821}, irrigating large tracts of land with multiple owners~\cite{ostrom_1990} and coordinating autonomous vehicles~\cite{Schwarting24972} are all vitally important mixed-motive problems, where individual incentives are not automatically aligned yet multi-agent cooperation is necessary to achieve beneficial outcomes for all. Often, such problems have a latent graph which structures the strategic interaction of the participants, such as the network cables linking computers, the irrigation channels supplying water, and the road system for vehicles. The field of network games~\cite{JACKSON201595} studies interactions between rational agents in the context of different topologies. However, network games abstract away crucial details informing the decision-making of agents, such as the geometry of the world and the temporal extent of strategic choices. 

Multi-agent deep reinforcement learning provides precisely the right tools to fill this gap in the literature, allowing us to simulate rational learned behavior as we vary properties of a spatio-temporally complex world. One approach to solving these problems is to invoke a single, centralized objective. However, this approach suffers from the ``lazy agent'' problem~\cite{DBLP:journals/corr/SunehagLGCZJLSL17}, where some agents learn to do nothing due to incorrect credit assignment. Moreover it does not generalize to situations involving interaction with humans. We must therefore look to uncover methods that incentivise cooperative behavior among rational self-interested agents. 

Game theory provides a precise framework for modelling the strategic choices of self-interested agents. There are two broad classes of interventions that promote cooperation: behavioural economics models of agent incentives (e.g. inequity aversion~\cite{10.2307/2586885}), and mechanism design for the game itself (e.g. auction design~\cite{10.2307/2696581}). Traditional game-theoretic models do not capture the rich cocktail of social-ecological factors that govern success or failure in real-world collective action problems, such as social preferences, geography, opportunity cost and learnability~\cite{Ostrom419}. With the advent of deep reinforcement learning, it has become possible to study these processes in detail~\cite{leibo2017multiagent, perolat2017multiagent}. Thus far, this work has mostly focused on the emergence of cooperation inspired by behavioral psychology~\cite{DBLP:journals/corr/abs-1810-08647, DBLP:journals/corr/abs-1803-08884, eccles2019learning} and algorithmic game theory~\cite{DBLP:journals/corr/LererP17, DBLP:journals/corr/abs-1709-04326, DBLP:journals/corr/abs-1902-03185}. We instead take a perspective motivated by mechanism design. 

In the most general sense, mechanism design examines how institutions can be constructed that result in better social behavior by groups of agents~\cite{jackson2014mechanism, doi:10.1080/19186444.2019.1591087}. Classically, a mechanism designer chooses a payoff structure to optimize some social objective. The payoff structure is known as a \textit{mechanism}. Network games in particular offer a rich space of design possibilities, by virtue of the underlying graph topology; see for example~\cite{AGARWAL2008520}. In spatio-temporally extended settings, we can naturally generalize the notion of a mechanism to generate a richer space, compromising not only interventions on the utility functions experienced by agents, but also subtle changes to the geographical layout and to the behavior of environmental components over time. Traditional analytic tools cannot simulate the effects of such mechanisms, for they reduce the complex environment to a more tractable abstract network game. Here we apply multi-agent deep reinforcement learning techniques to perform a simulation in a complex gridworld, and explore the effect of a wide range of novel manipulations. This opens the question of automatically selecting mechanisms for particular objectives, which we leave for further study. 

\textbf{Our contribution}. The aims of this paper are (1) to introduce a new method for modelling the behavior of self-interested agents in collective action problems with topological, geometrical and temporal structure, and (2) to use this method to draw conclusions relevant for mechanism design that promotes cooperative behavior. Our method comprises a rigorous protocol for intervening on a spatio-temporally complex environment, and modelling the effects on social outcomes for rational agents via multi-agent deep reinforcement learning. To illustrate this general method, we introduce a new gridworld domain called \textit{Supply Chain}, in which agents are rewarded for processing goods according to a given network structure. The actions of agents are interdependent: downstream agents rely on upstream agents to deliver goods to process. Furthermore, processing centres require periodic maintenance from more than one agent. This yields a collective action problem with latent graph structure: each agent obtains individual reward by processing items, but the public good is best served when the agents occasionally cooperate to keep the supply chain intact. 

For environmental interventions, we take the perspective of a system designer and ask: what mechanisms might we introduce to the world, and how do these affect the cooperation of agents? We not only vary the topology of the world, as in traditional network games, but also the geometry, maintenance cost, and agent specialization. In all cases we find an intricate interplay between incentive structure, multi-agent interaction and learnability which affects the nature of emergent cooperation. More precisely, we introduce a metric of care in order to understand these dynamics~\cite{dc9fd7d5c9714e7da5dbe5f0f855a767}. We find that reciprocal care is diminished when the maintenance burden is lower, and that reciprocity is promoted by training generalist agents that can operate any station in the supply chain, rather than specialists. We do not expect the conclusions we draw to have general applicability; rather we argue that this case study demonstrates the power and insight provided by our new method. 

\subsection{Related work}

We are not the first to apply machine learning techniques in problems related to mechanism design. In auctions, the simulation and optimization of mechanisms has been explicitly performed via a variety of learning methods~\cite{conitzer2002complexity, DBLP:journals/corr/abs-1907-05181, 1530752, DBLP:journals/corr/DuttingFNP17, tang2017reinforcement}. In the context of matrix game social dilemmas,~\cite{DBLP:journals/corr/abs-1806-04067} used multi-agent reinforcement learning to adaptively learn reward-giving mechanisms that promote cooperative behavior. \cite{zheng2020ai} also takes the perspective of a social planner to directly optimize taxation policies for agents in a gridworld setting. Multi-agent reinforment learning in particular has been applied to design mechanisms for abstract social networks \cite{ahlgren2020wes}. Our work builds on this literature, extending the notion of a mechanism to include a wide variety of interventions that inform decision-making under learning: geometrical, topological and temporal. By virtue of this added complexity, we focus here on the simulation of learning agents under different mechanistic interventions, rather than the optimization of the mechanisms themselves. 

There also is an extensive literature on multi-agent learning in supply chain problems. The archetypal abstraction is provided by the Beer game~\cite{forrester1958industrial}, an extensive form game that captures some dynamics of supply chain coordination. In~\cite{DBLP:journals/corr/abs-1708-05924}, the authors treated the Beer game as a fully cooperative problem, and simulated the behavior of deep Q learners. Similarly~\cite{pontrandolfo} and~\cite{kemmer} investigate how reinforcement learning agents perform in network games that approximate supply chain dynamics. Our work is very much in the same spirit, albeit with three important differences. Firstly, we treat the supply chain as a collective action problem rather than as a fully cooperative one. As such, the emergence of cooperation is not a given, rather it is heavily influenced by the environmental and learning dynamics, as we shall see in Section \ref{sec:results}. Secondly, our supply chain is embedded in a complex gridworld, allowing us to study the emergence of cooperation in great detail. Finally, our methods are sufficiently general to apply to other situations with underlying graph structure; the supply chain is a first example, rather than the primary application. 

\begin{figure*}[h]
    \begin{subfigure}{0.6\textwidth}
        \centering
        \includegraphics[height=11em]{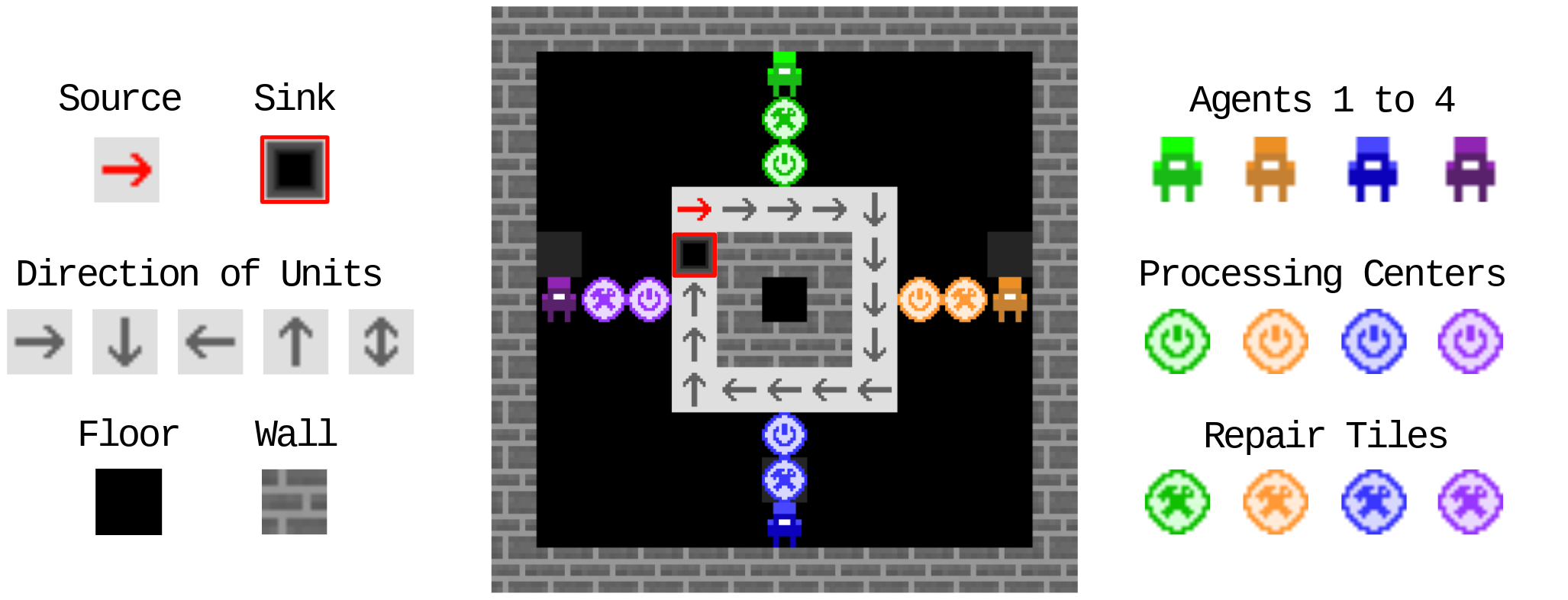}
        \caption{The state of the environment at the start of an episode.}
    \end{subfigure}
    \hfill
    \begin{subfigure}{0.35\textwidth}
        \centering
        \includegraphics[height=11em]{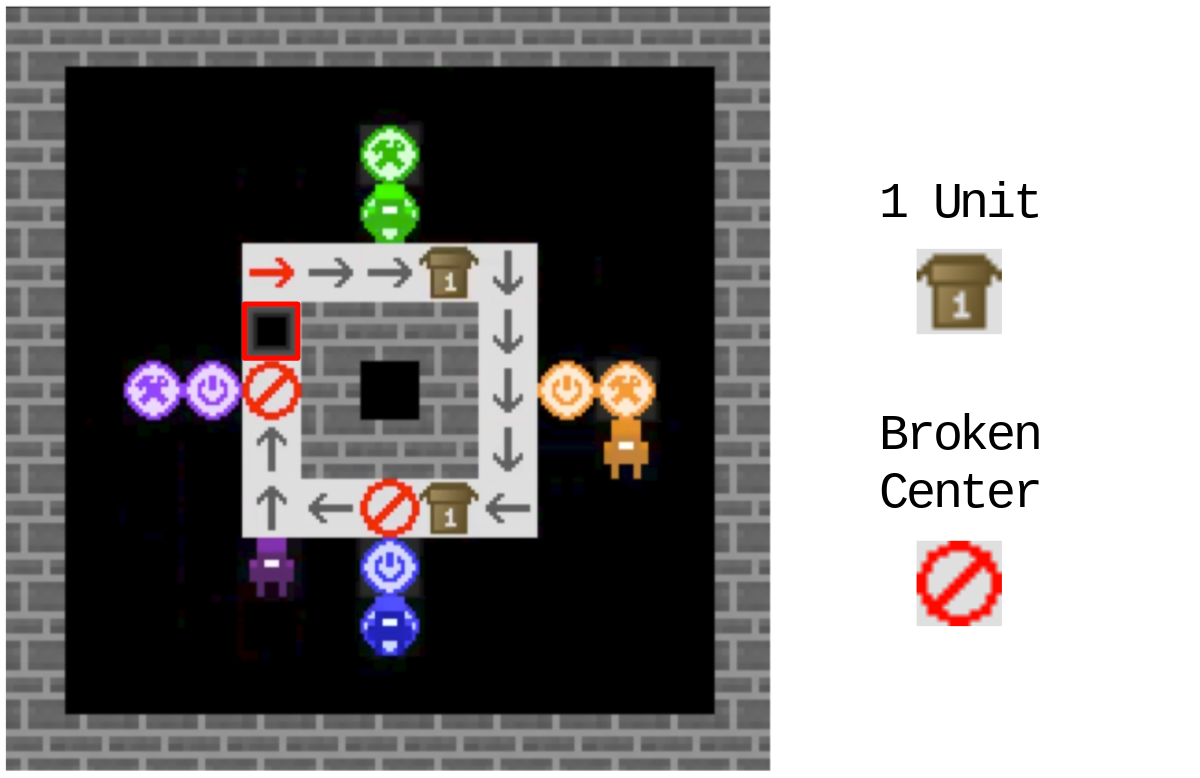}
        \caption{An example state mid-episode.}
    \end{subfigure}
    \caption{The Supply Chain environment, visualized with a circular layout. The sprite representations in the figure are for clarity of human interpretation. Agents instead receive 13x13 pixels observations where each type of object has as a unique RGB color. }
    \label{fig:environment}
\end{figure*}

\section{Methods}

\subsection{The Supply Chain environment}

The Supply Chain environment (Figure~\ref{fig:environment}) is a $2$-dimensional gridworld in which agents must maintain their own individual \textit{processing centers} with the help of other agents, in order to process \textit{units} passing through the supply chain. These units enter the environment via the source tile and travel through the supply chain until they reach the sink tile where they are removed. Importantly, units stop next to each processing center and do not continue along the supply chain until they have been processed by that processing center's owner. This is achieved by the owning agent standing on their associated processing center tile, thereby processing the unit which allows it to continue and giving $+1$ reward to that agent.

Upon processing a unit, there is a 25\% chance that the agent's processing center breaks down. This stops units from passing through the supply chain. There are two mechanisms by which a processing center can be repaired: 
\begin{enumerate}
    \item Automatically with probability $p=\frac{1}{\textrm{repair time}}$ each step (referred to as ``self-repair'').
    \item Manually by two agents standing on both the processing center tile and associated repair tile (referred to as ``two-agent repair'').
\end{enumerate} With the exception of the experiments performed in Section~\ref{sec:self_repair}, we disable self-repair and consider only two-agent repair.

An episode in this environment lasts $1000$ steps. On each step, there is a 10\% chance of a unit entering the environment at each source tile, and all other units in the supply chain move once if the next space in the supply chain is unoccupied. If the next space is occupied by another unit, the moving unit is permanently discarded from the supply chain and therefore cannot be processed by any agent, leading to a lost opportunity to obtain reward.

This environment is a directed graph-structured collective action problem. The supply chain on which items flow is the spatio-temporal realization of an abstract network game. Collective action is required to maintain the processing centers. In particular, each agent would prefer that others took on the responsibility for fixing broken processing centers, since this comes at the opportunity cost of processing units themselves. However, if all agents refuse to cooperate, then they receive low group reward, since the processing centers break, and no units can be processed. 

The spatio-temporal nature of the environment and the underlying topology of the supply chain admit a wide range of mechanistic manipulations. These manipulations interact with agent learning in intricate ways, significantly altering the equilibria at convergence. More precisely, depending on the environment properties we see different patterns of ``care'' between agents, understood in terms of help provided to others when repairing broken processing centers. In particular, under certain circumstances reciprocity~\cite{chammah1965prisoner, Axelrod1390, trivers71} may arise naturally, promoted by the underlying graph. In Section 3, we manipulate the auto-repair of processing centers, the geometry of the processing center layout, and the topology of the supply chain, drawing conclusions about which mechanisms promote and suppress the emergence of care.  Unless specified otherwise, the experiments are performed using the environment in Figure\,\ref{fig:environment}.

\subsection{Social outcome metrics}

Consider an instance of the \textit{Supply Chain} game with episode length $T$ and $N$ agents, uniquely assigned to $N$ processing centers such that agent $i$ always processes units at center $i$. Each supply chain has an underlying directed graph structure $G = (V,E)$ with a set $V$ of $N$ vertices, and a set $E$ of ordered pairs of centers which are the directed edges that make up the supply chain itself. Each vertex without incoming edges is a center that receives units from a source tile while each vertex without outgoing edges is a center from which units flow towards a sink tile. For each center, we use $\textsf{UP}_G(i)$ to denote the set of vertices that are upstream from $i$, i.e. from which a unit could flow to $i$, while $\textsf{DO}_G(i)$ denotes the set of centers in $G$ that are downstream from $i$. 
Let $r_t^i$, $b_t^{i}$ and $c_t^{ij}$, be binary variables that specify if, at time $t$, agent $i$ respectively processes a unit, breaks its processing center, or repairs a processing center of agent $j$.
In turn, $R^i$, $B^i$, and $C^{ij}$ are the number of processed units (reward), breakages, and repairs aggregated over one episode.

Building on these definitions, we introduce a set of metrics that help analyze social outcomes:
\begin{itemize}
    \item The \emph{care matrix} ($C$) with elements $C^{ij}$ tracks the care (repairs) each agent has received from each other agent, relative to the total number of breakages $\sum_i B^i$. 
    \item The \emph{care reciprocity} ($S$) measures how symmetric the care matrix is
    \begin{equation}
        S = \frac{\|C_{sym}\|_2-\|C_{anti}\|_2}{\|C_{sym}\|_2+\|C_{anti}\|_2}
    \end{equation}
    where $C_{sym}=\frac{1}{2}(C+C^\mathsf{T})$ and $C_{anti}=\frac{1}{2}(C-C^\mathsf{T})$. $S=1$ means the matrix is symmetric and $S=-1$ that the matrix is anti-symmetric (note that because $C$ is non-negative, $S \ge 0$). 
    \item The \emph{average care direction} ($D$) measures whether care is entirely upstream ($D=1$), entirely downstream ($D=-1$) or in between ($-1<D<1$), 
    \begin{equation}
        D = \frac{1}{\sum_{i=1}^N  \sum_{j=1}^N C^{ij}}\sum_{i=1}^N  \sum_{j=1}^N \left([j \in \mathsf{UP}_G(i)] - [j \in \mathsf{DO}_G(i)]\right)C^{ij}.
    \end{equation}

\end{itemize}

\subsection{Multi-Agent Reinforcement Learning}
We train agents using the advantage actor-critic (A3C)~\cite{mnih2016asynchronous} learning algorithm using 400 parallel environments to generate experience for each learner. Episodes contain 4 agents which are sampled without replacement from a population of 8 and assigned to random processing centers in the environment~\cite{DBLP:journals/corr/abs-1807-01281}\footnote{We also run an experiment where agents are instead to the same processing center for the course of training, see Section \ref{sec:spec}}. Every agent uses their own neural network and is trained for $10^9$ steps by receiving importance-weighted policy updates~\cite{espeholt2018impala}.

\subsection{Architecture}
For each observation, consisting of 13x13 RGB pixels, the neural network architecture computes a policy (probability per action) and value (for each observation). It consists of a visual encoder, which projects to a 2-layer fully connected network, followed by an LSTM, which in turn projects via another linear map to the policy and value outputs. The visual encoder is a 2D-convolutional neural net with one layer of 6 channel with a kernel and stride size of 1. The fully connected network has 64 ReLU neurons in each layer. The LSTM has 128 units. The action space corresponds to five actions: moving up, moving down, moving left, moving right, or wait one timestep. The actions themselves naturally generalize to other gridworld environments.

\subsection{Training parameters}

For agent learning, we use a discount-factor of $0.99$, a batch size of $16$, an unroll length of $100$ for backpropagation through time, while the weight of entropy regularisation of the policy logits is $0.003$. We use the RMS-prop optimizer with learning rate $0.0004$, epsilon $10^{-5}$, momentum $0.0$, and decay $0.99$. The agent also minimizes a contrastive predictive coding loss in the manner of an auxiliary objective.

\begin{figure*}[h]
    \begin{subfigure}{0.32\textwidth}
        \centering
        \includegraphics[width=\linewidth]{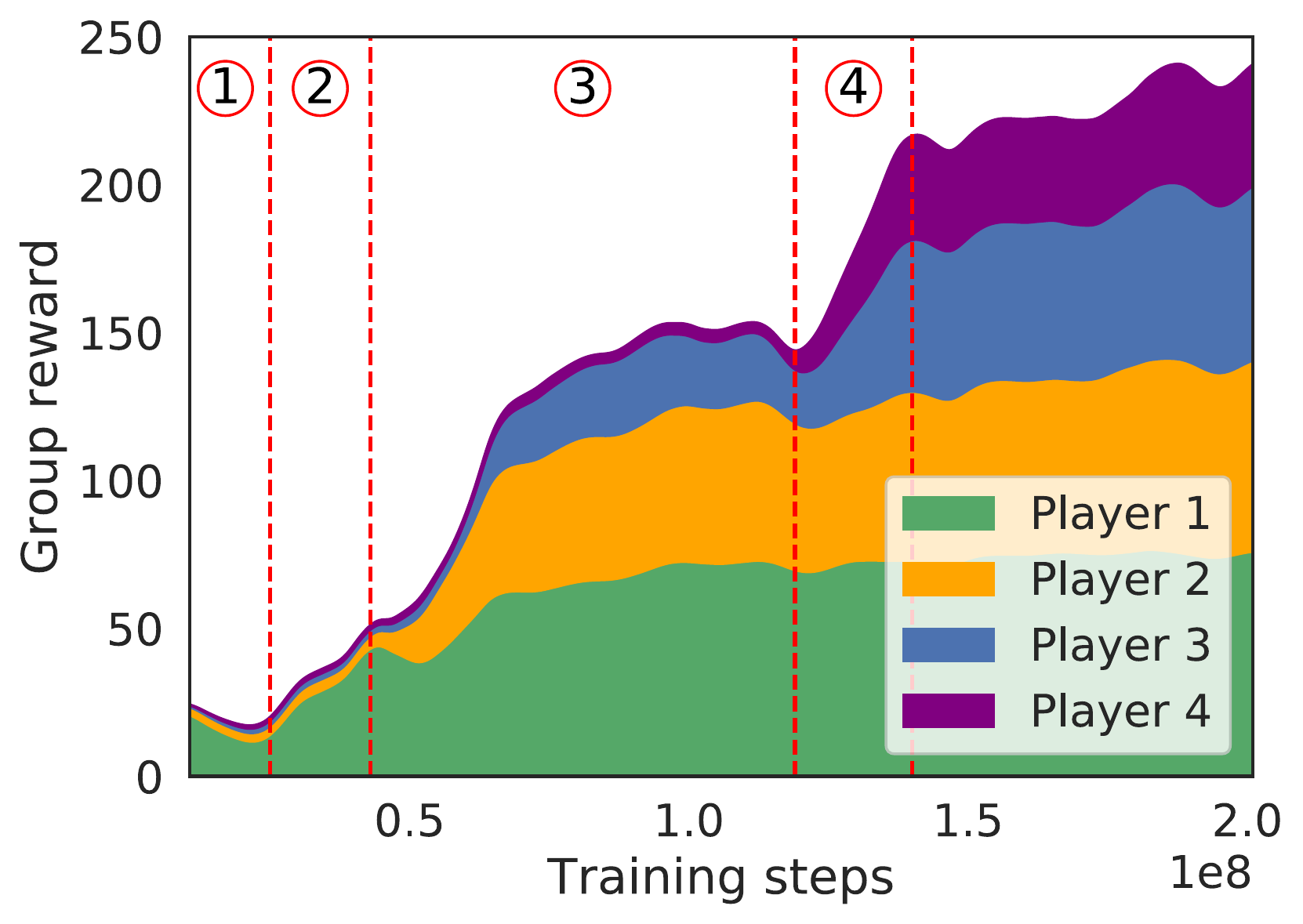}
        \caption{Individual reward per agent stacked for group reward.}
    \end{subfigure}
    \;
    \begin{subfigure}{0.32\textwidth}
        \centering
        \includegraphics[width=\linewidth]{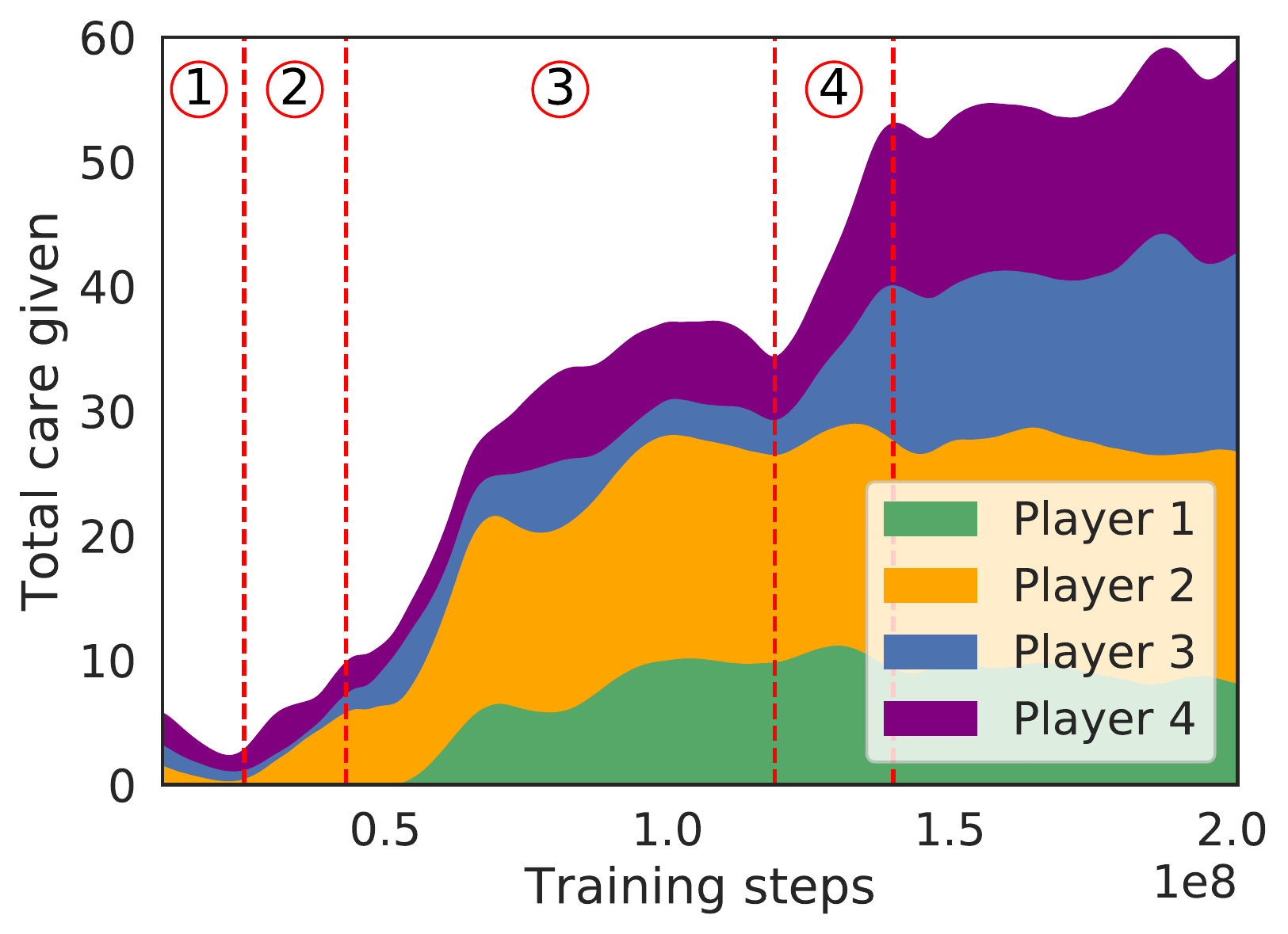}
        \caption{Care given per agent stacked for total care given.}
    \end{subfigure}
    \;
    \begin{subfigure}{0.32\textwidth}
        \centering
        \includegraphics[width=\linewidth]{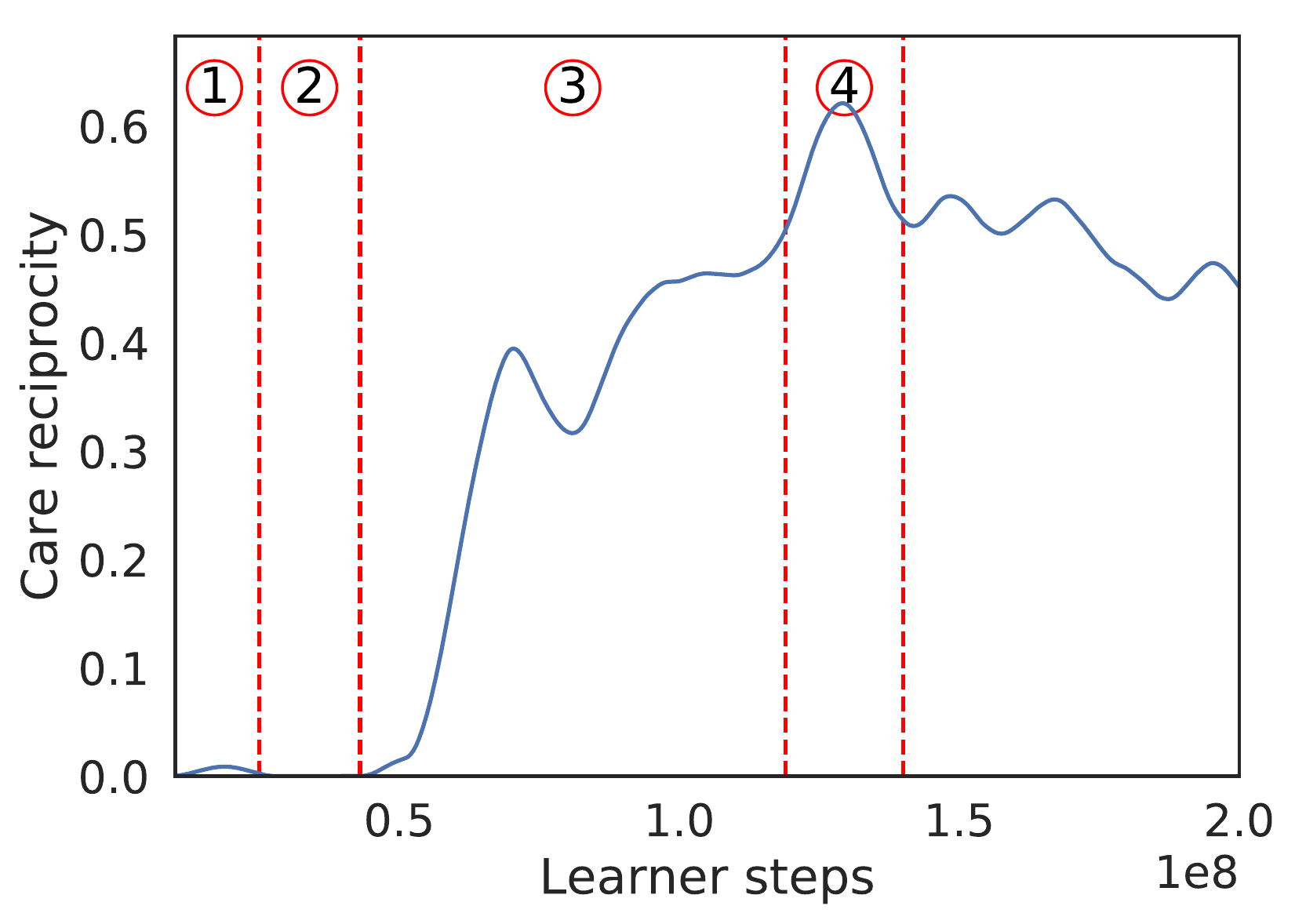}
        \caption{Care reciprocity.}
    \end{subfigure}
    \caption{Evolution of social outcome metrics over the course of training on a circular chain with four agents and self-repair disabled. Please see Figure  \ref{fig:learning_curves_carematrices} for the care matrices at the end of each phase.}
    \label{fig:learning_curves_reciprocity}
\end{figure*}

\begin{figure*}[h]
    \begin{subfigure}{0.24\textwidth}
        \centering
        \includegraphics[width=\linewidth]{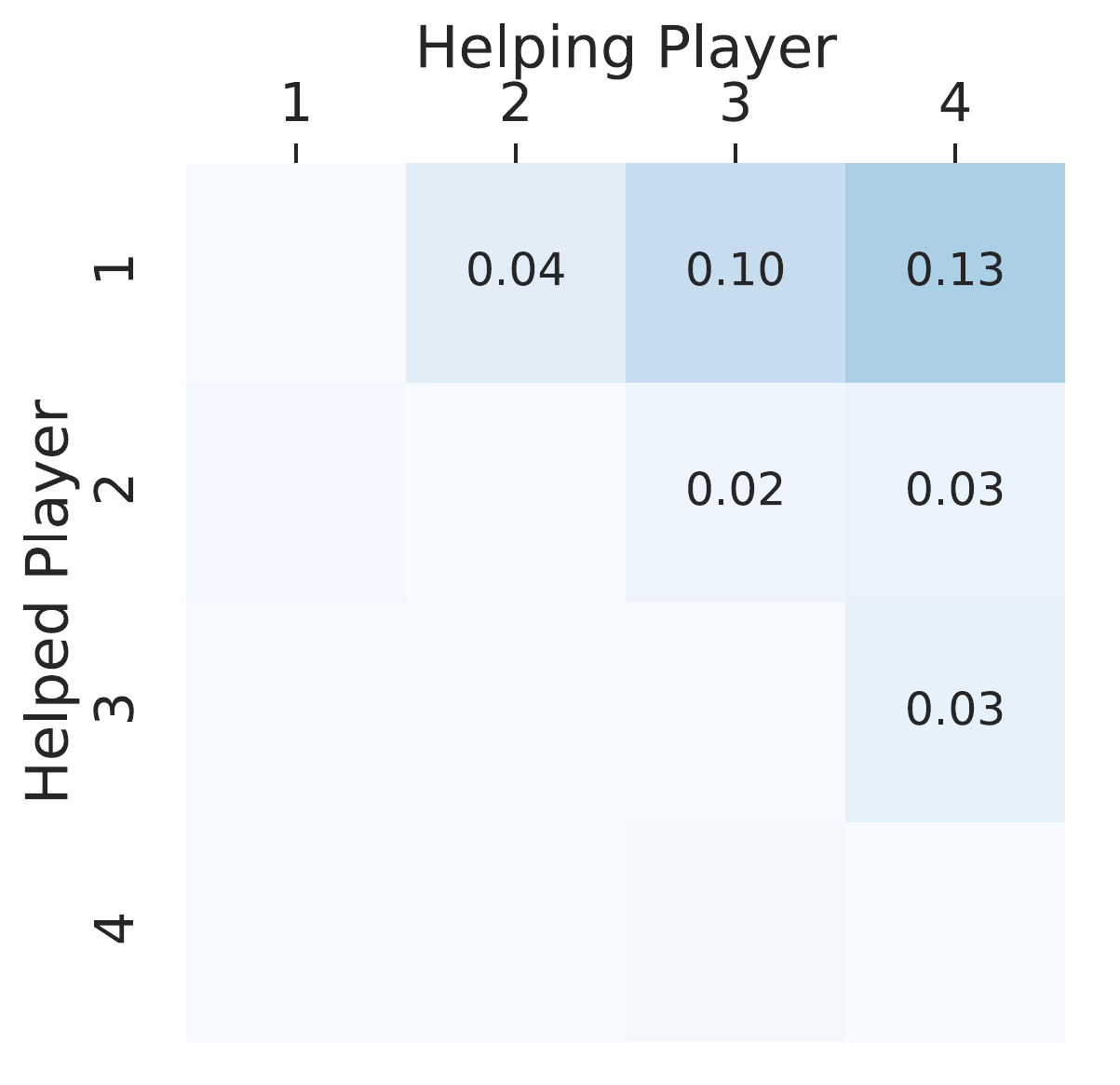}
        \caption{End of phase 1.}
    \end{subfigure}\;
    \begin{subfigure}{0.24\textwidth}
        \centering
        \includegraphics[width=\linewidth]{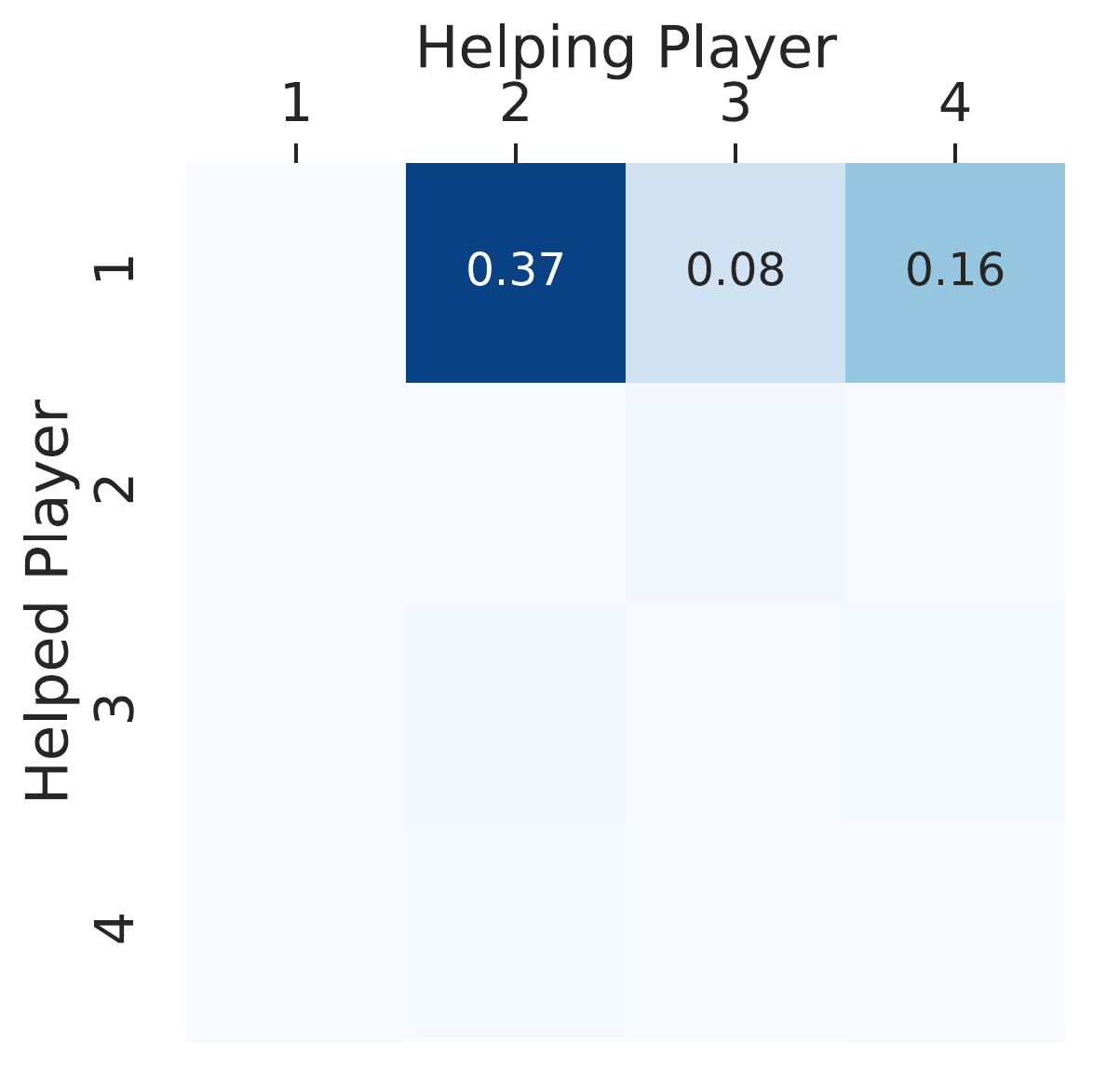}
        \caption{End of phase 2.}
    \end{subfigure}\;
    \begin{subfigure}{0.24\textwidth}
        \centering
        \includegraphics[width=\linewidth]{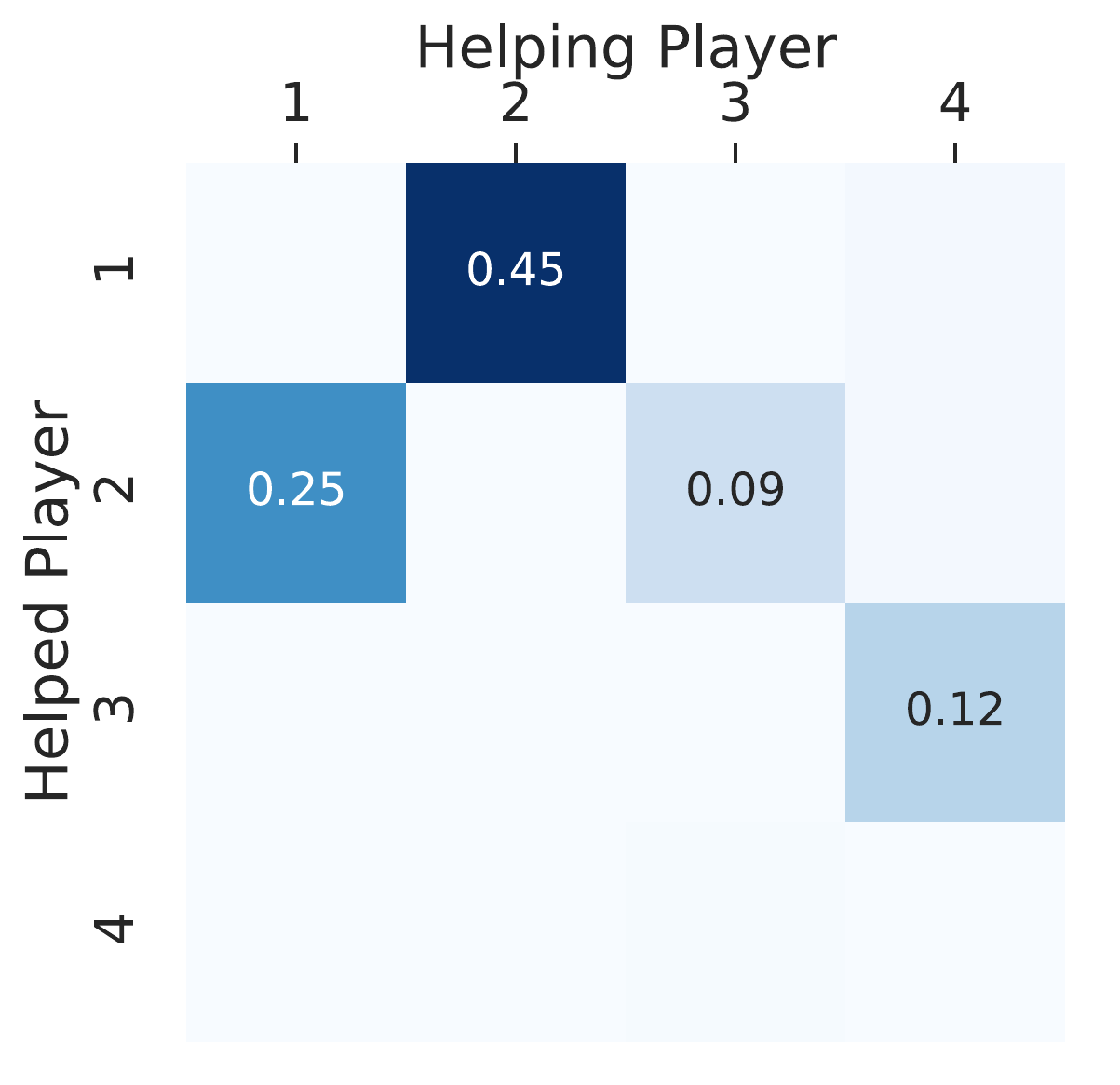}
        \caption{End of phase 3.}
    \end{subfigure}\;
    \begin{subfigure}{0.24\textwidth}
        \centering
        \includegraphics[width=\linewidth]{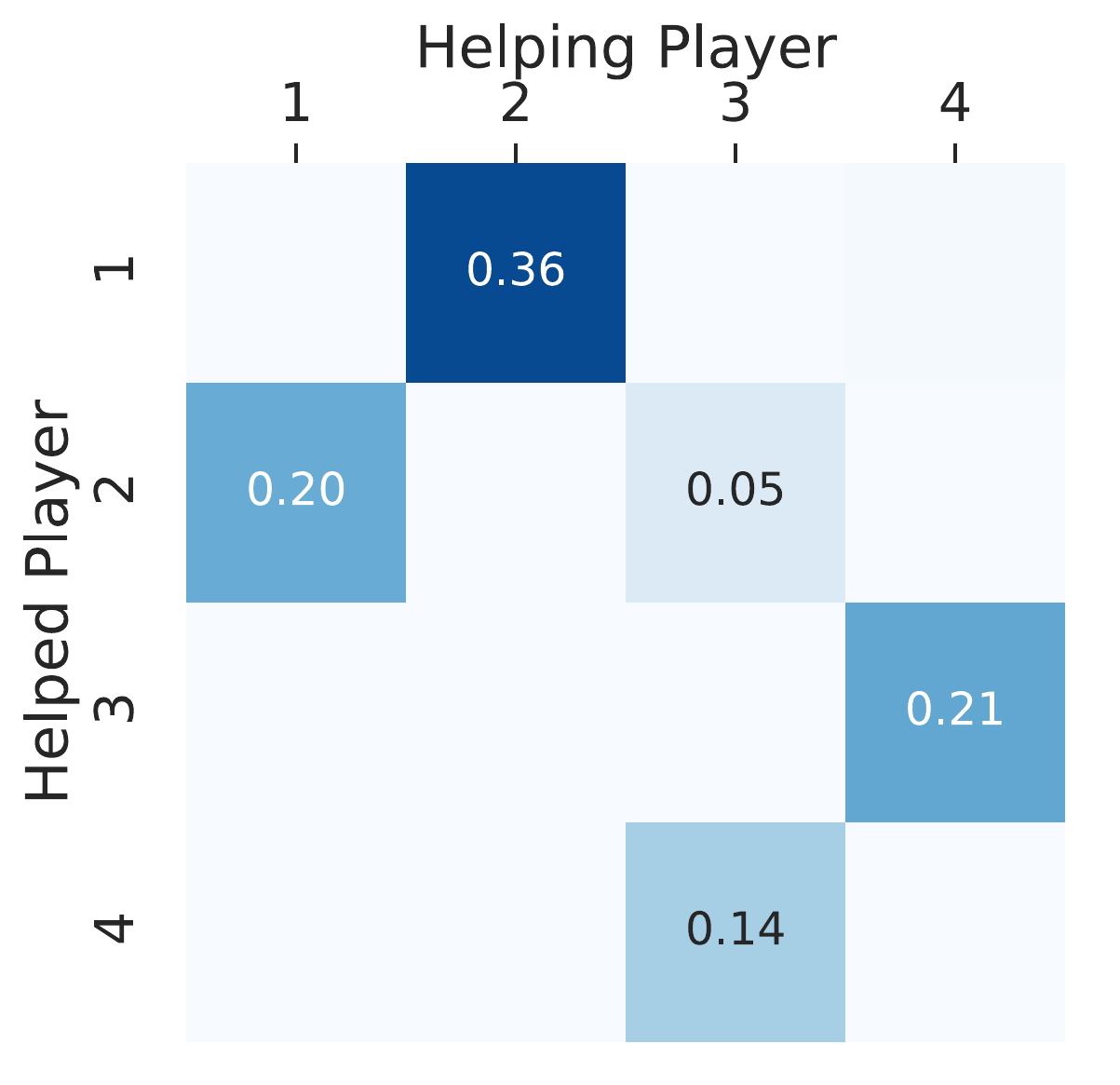}
        \caption{End of phase 4.}\label{fig:baseline}
    \end{subfigure}
    \caption{Care matrices at the end of the four distinct learning phases. For improved readability, values below 0.01 are omitted.}
    \label{fig:learning_curves_carematrices}
\end{figure*}

\section{Results}\label{sec:results}
In this section, we study how different environmental interventions influence social outcomes. In the Supply Chain environment, we analyze the learning dynamics and the emergence of reciprocal care. At convergence, we study the effect of changing if and how fast processing centers can repair themselves autonomously, and we increase the inter-center distance to study the influence of geometry. Finally, we discuss how subtle changes in the environment's underlying graph structure can drastically change social outcomes. Unless specified otherwise, the experiments are performed using the environment in Figure\,\ref{fig:environment}.

\subsection{Learning dynamics and the emergence of care}\label{sec:emergence}

\begin{figure*}[h]
\centering
    \begin{subfigure}{0.32\textwidth}
        \centering
        \includegraphics[width=\linewidth]{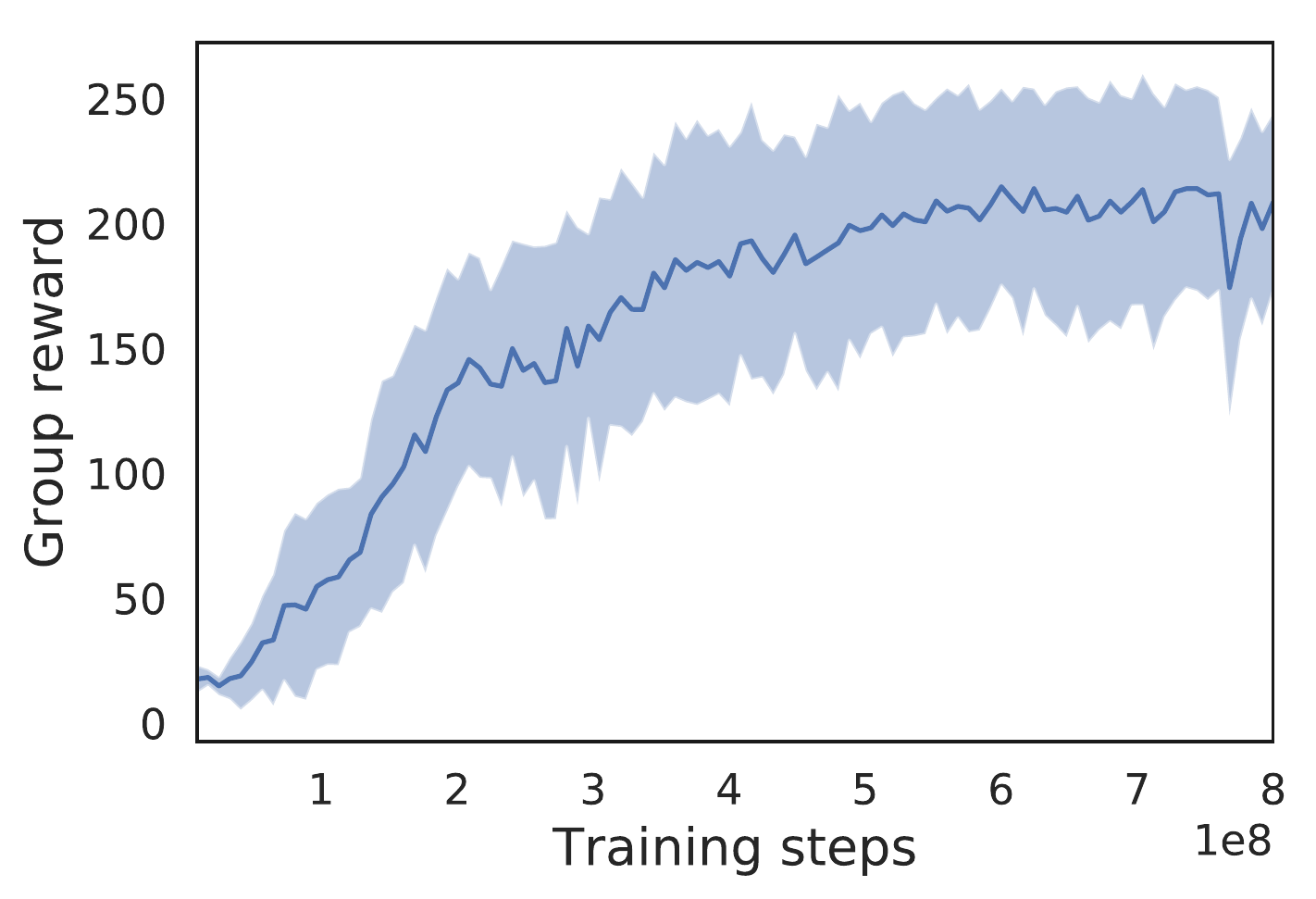}
        \caption{Group reward.}
    \end{subfigure}
    \begin{subfigure}{0.32\textwidth}
        \centering
        \includegraphics[width=\linewidth]{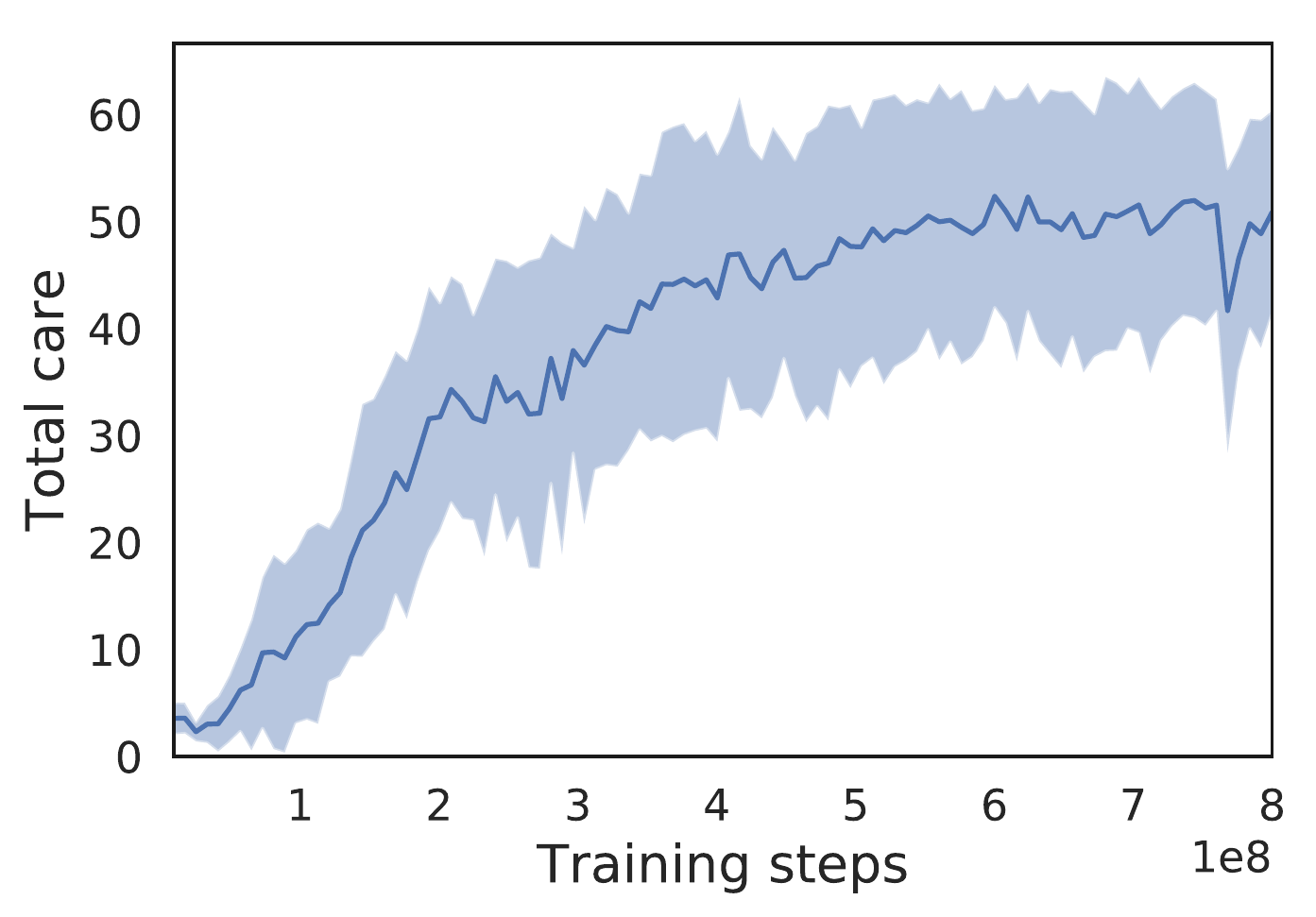}
        \caption{Total care.}
    \end{subfigure}\;
    \begin{subfigure}{0.32\textwidth}
        \centering
        \includegraphics[width=\linewidth]{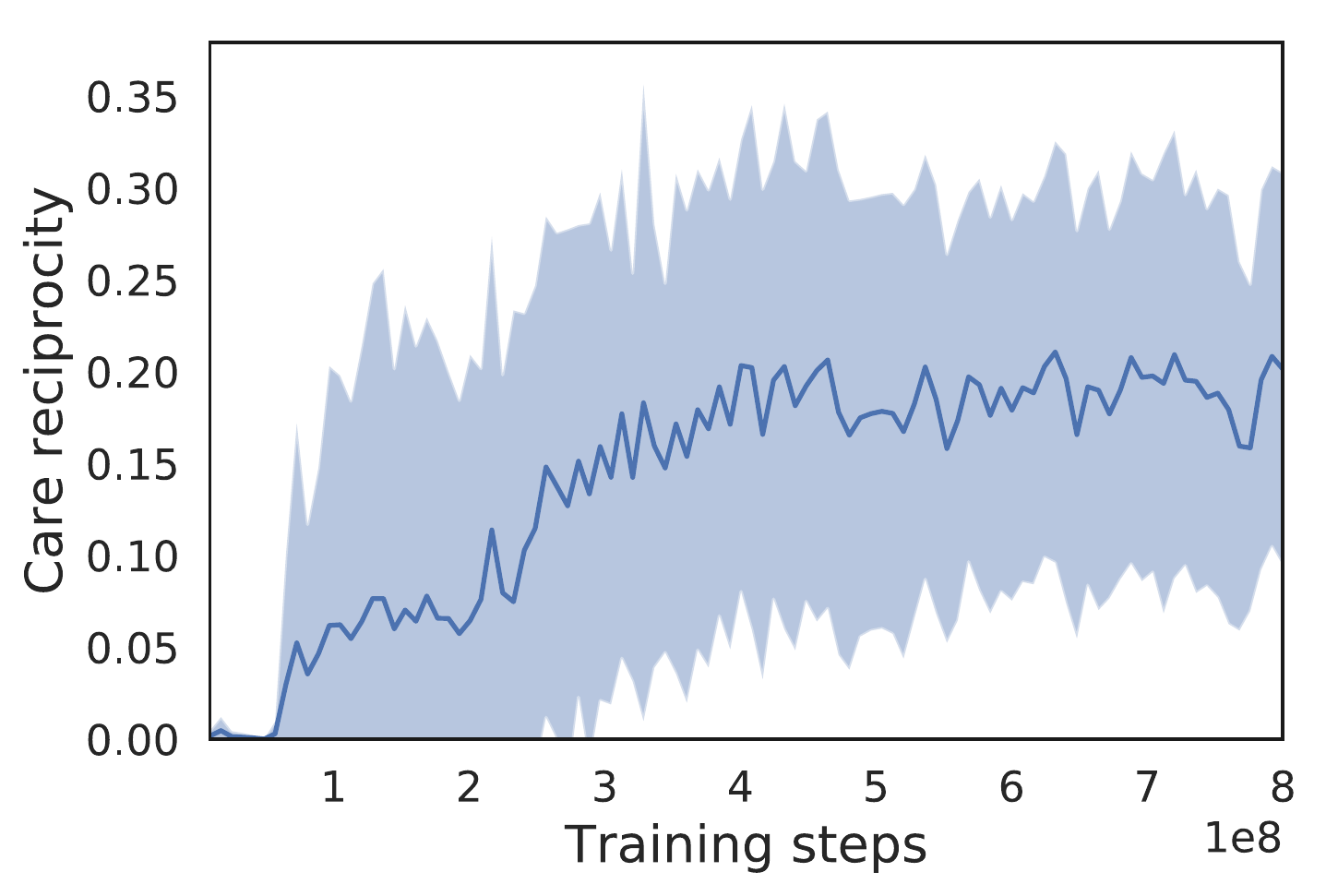}
    \caption{Care reciprocity.}
    \end{subfigure}\;
    \caption{Evolution of different social outcome metrics over the coarse of training, averaged across 8 runs with different random seeds. The shaded regions represent $95\%$ confidence intervals.}
    \label{fig:averagedlearningcurves}
\end{figure*}

We first consider the case in which self-repair has been turned off and processing centers can only be repaired by other agents. Agents need to cooperate with other agents to ensure their processing centers are repaired. However, learning to care for others only benefits agents indirectly and is thus a more complex behavior to learn than simply processing units. Learning to care happens in distinct phases, each characterized by different behaviors and social outcomes. Each phase features a rather abrupt change in the individual reward received: once an agent has found an improved strategy via exploration, it is quickly able to exploit this, shifting the equilibrium dramatically. Accordingly, each phase can last shorter or longer depending on when the agents ``discover'' the new behavior.
We therefore analyze the outcomes after each phase, similarly to~\cite{baker2019emergent,perolat2017multiagent}, and fist analyze one typical run before analyzing the averages across mutliple runs. A second typical run can be found in the Appendix. The behaviors observed in these two runs are archetypal for the behaviors that are observed in all eight runs.

In Figure \ref{fig:learning_curves_reciprocity}, we observe the individual rewards, the total (unnormalized) care per agent $C^i=\sum_j C^{ij}$, and the care reciprocity. The dashed lines represent the phase transitions between $4$ distinct training phases until convergence. The full care matrix after each phase can be found in Figure \ref{fig:learning_curves_carematrices}. Phase 1 begins at the start of training and ends after $\approx 2\cdot 10^7$ training steps. It is characterized by agents learning how to navigate the environment, process units and explore upstream caring (indicated by the non-zero values in the upper-right triangle of \ref{fig:learning_curves_carematrices}a). Agents have yet to learn when repairing is beneficial, and we thus observe the highest average reward for the agent that is closest to the source and a progressively lower reward for agents further downstream, since units can only reach these agents if all centers upstream are not broken. Phase 2 ends after $\approx 4\cdot 10^7$ steps and is characterized by the first examples of consistent care. The second agent (and to a lesser extent the third and fourth) learn that they can earn more reward by repairing the first processing center. Phase 3 ends after $\approx 1.2\cdot 10^8$ steps. In this phase, agent 2 learns that repairing center 1 only results in more reward when it can process units at its own center. To keep this incentive for agent 2, agent 1 thus learns to reciprocate the care received by agent 2. This leads to a sudden increase in care reciprocity and a drastic increase in reward for agent 2. At the same time, agent 4 learns to help agent 3 in order to receive more reward, similar to what we observed for the first two agents in phase 2. Finally, during phase 4, reciprocity emerges between agents 3 and 4 as agent 3 learns that agent 4 only repairs when its own processing center is fixed. Only in phase 4 are there an appreciable number of units delivered to the sink: note that this quantity is equal to the reward received by agent 4. 

In Figure \ref{fig:averagedlearningcurves}, we show the same social outcome metrics but now averaged across $8$ runs. Due to the abrupt nature of the learning process, we no longer observe the clear phase transitions that we observed in a single run and observe a high relative variance for all the metrics. The latter is especially clear or the care reciprocity as the the amount of reciprocal care does not only differ during learning but also at convergence. Naturally, as receiving care is necessary to earn any additional reward beyond the trivial reward that can be earned before the first processing centers break, we observe a strong correlation between the group reward and the total care. 

In a sense, it is surprising that reciprocity emerges in this environment. Na\"ively, one might expect that the collective action problem is dominated by selfish incentives, and an explicit model-based intervention may be required to solve the social dilemma, as in~\cite{eccles2019learning, pey11}. However, in our case, the underlying graph structure organizes agent interactions in such a way as to promote the emergence of reciprocity. This is exactly in line with previous work in abstract network games~\cite{ohtsuki2007direct}, extending it to a setting where we can examine mechanism design in detail. 

\begin{figure}
    \centering

    \includegraphics[width=0.24\textwidth]{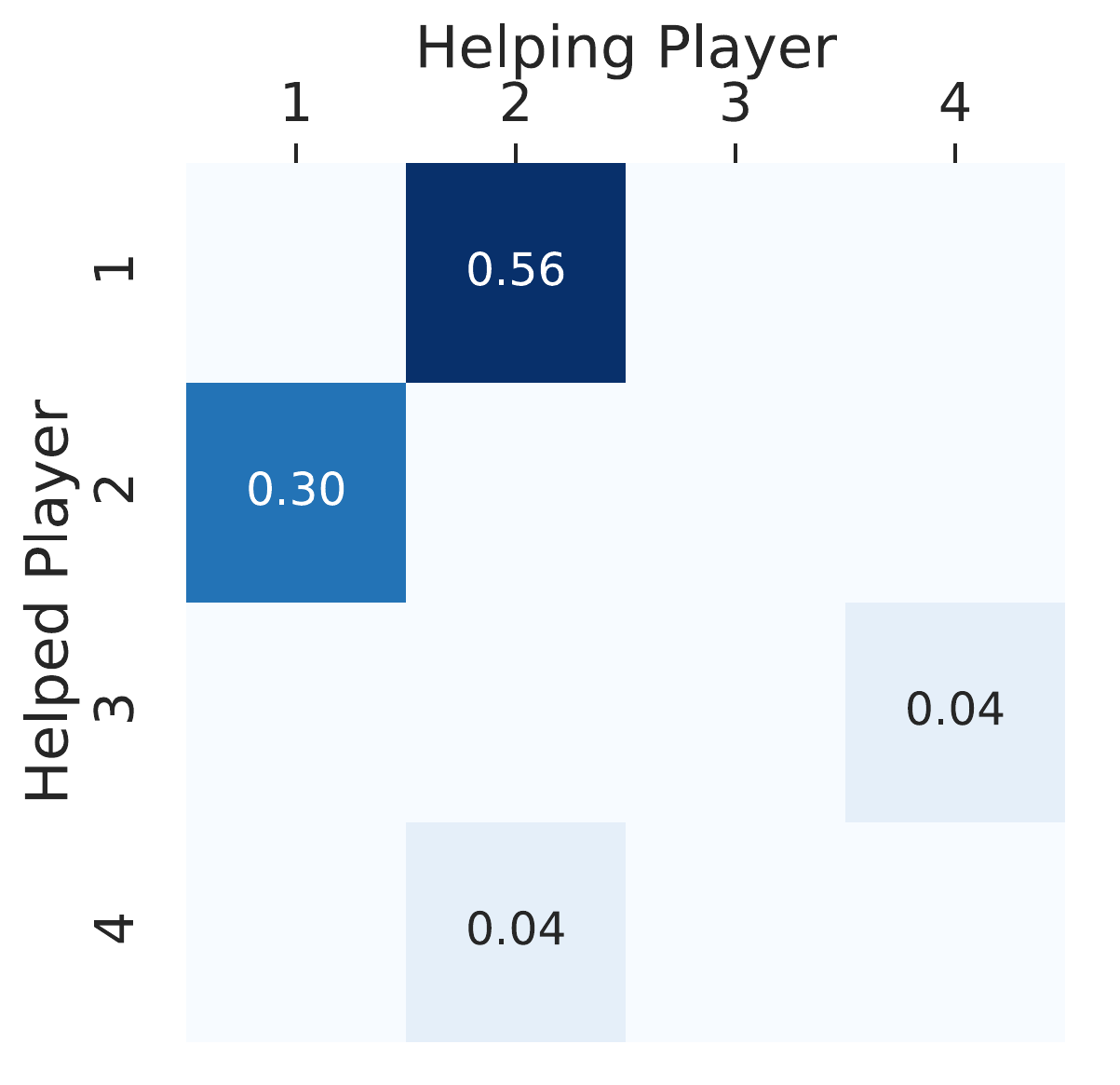}

    \caption{Reciprocal care for agent 4 is not observed when agent 3 is replaced by a selfish agent at test time. Compare the care matrix with Figure \ref{fig:baseline} for an evaluation without replacing agent 3 for a `selfish' agent.}
    \label{fig:selfishwithnonselfish}
\end{figure}

Reciprocity, in the sense of tit-for-tat, requires that I repair your processing center if and only if you do the same for me. This is a complex, conditional and fundamentally temporal strategy. We perform an experiment to check that our agents have learned reciprocity in this strong sense. In Figure \ref{fig:selfishwithnonselfish}b, we examine the care matrix while evaluating, after convergence, three `caring' agents and one `selfish' agent in a single environment. The caring agents, assigned to centers 1, 2 and 4, were trained in an environment where self-repair is turned off and agents thus learn to care while agent 3 was trained in environment with fast self-repairing (repair time $=10$) and thus has never learned to care for other agents (see Section~\ref{sec:self_repair}). Interestingly, not only does selfish agent 3 stop caring for agent 4 but, without the care being reciprocated, agent 4 also stops caring for agent 3. We find that, instead of agents learning to care unconditionally, they learn that giving care is only beneficial if its reciprocated and thus if their own processing center is also being cared for.

\begin{figure*}[ht!]
    \begin{subfigure}{0.32\textwidth}
        \centering
        \includegraphics[width=0.95\linewidth]{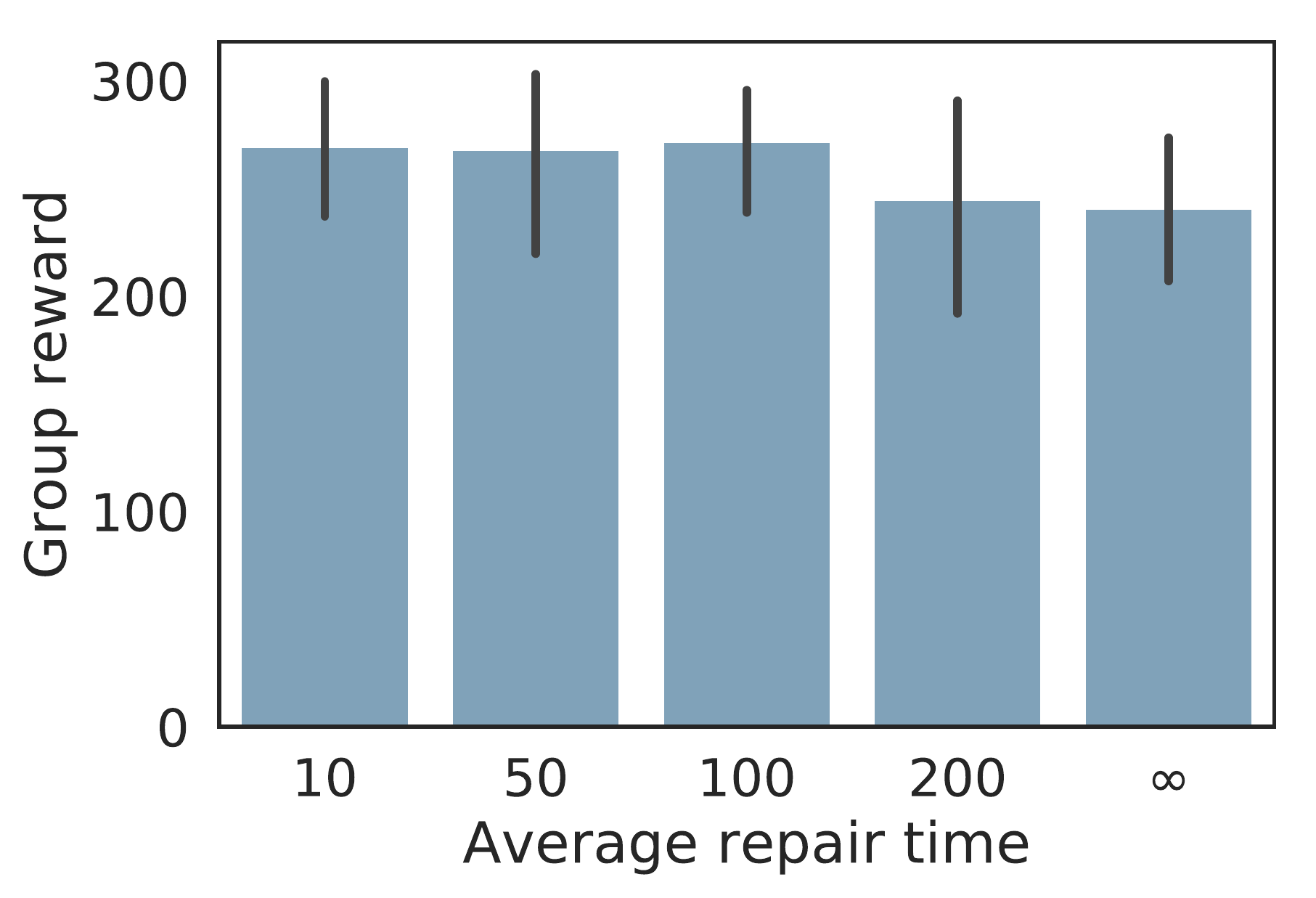}
        \caption{Group reward.}
    \end{subfigure}
    \;
    \begin{subfigure}{0.32\textwidth}
        \centering
        \includegraphics[width=0.95\linewidth]{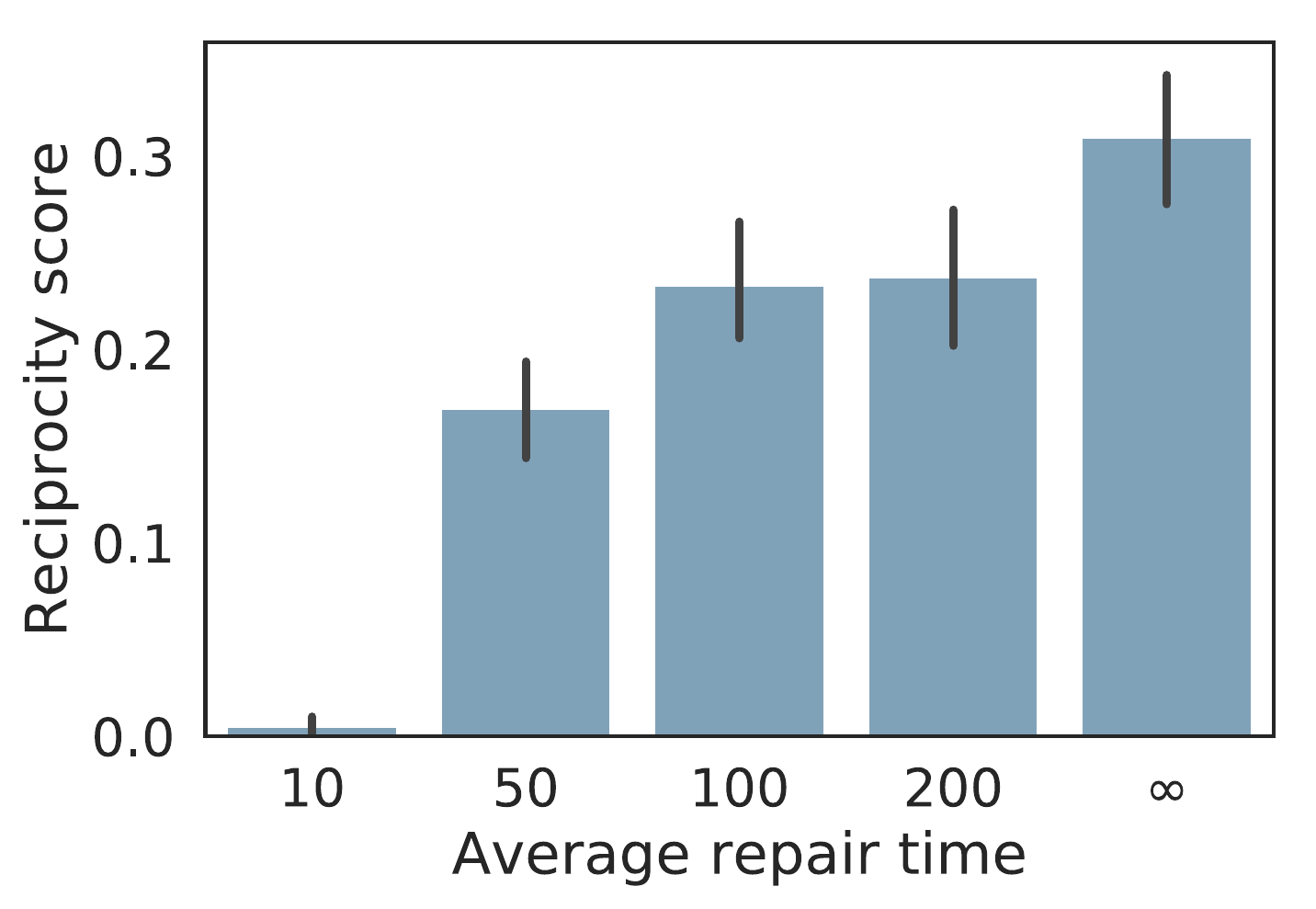}
        \caption{Care reciprocity.}
    \end{subfigure}
    \;
    \begin{subfigure}{0.32\textwidth}
        \centering
        \includegraphics[width=0.95\linewidth]{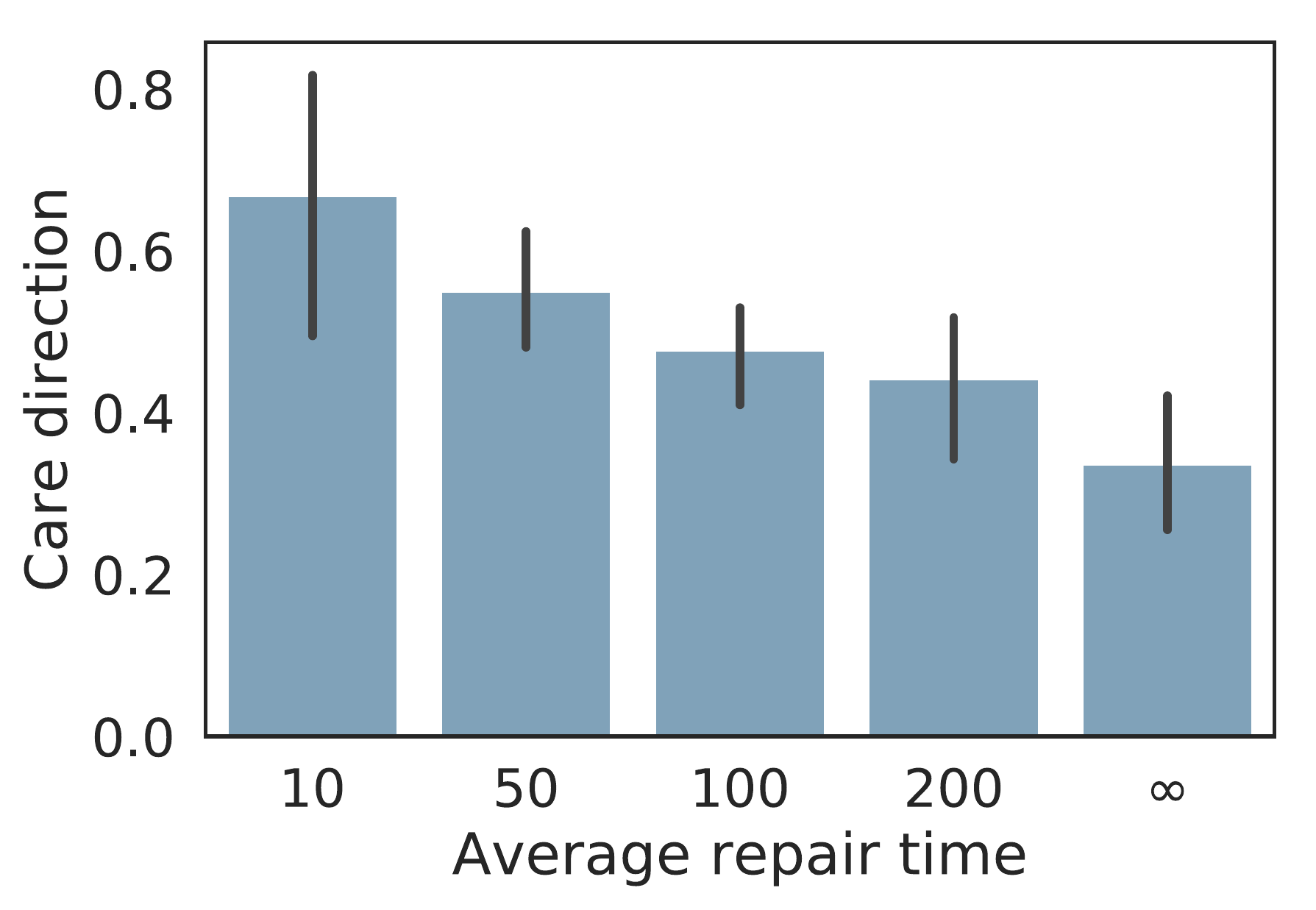}
        \caption{Care direction.}\label{fig:selfrepdirection}
    \end{subfigure}
    \caption{Social outcome metrics for different values of the repair time averaged over $8$ different runs where repair time $ =\infty$ denotes that self-repair has been disabled.}
    \label{fig:evalutate_averagerepairtime}
\end{figure*}

\begin{figure*}[h!]
    \centering
    \begin{subfigure}{0.32\textwidth}
        \centering
        \includegraphics[width=0.95\linewidth]{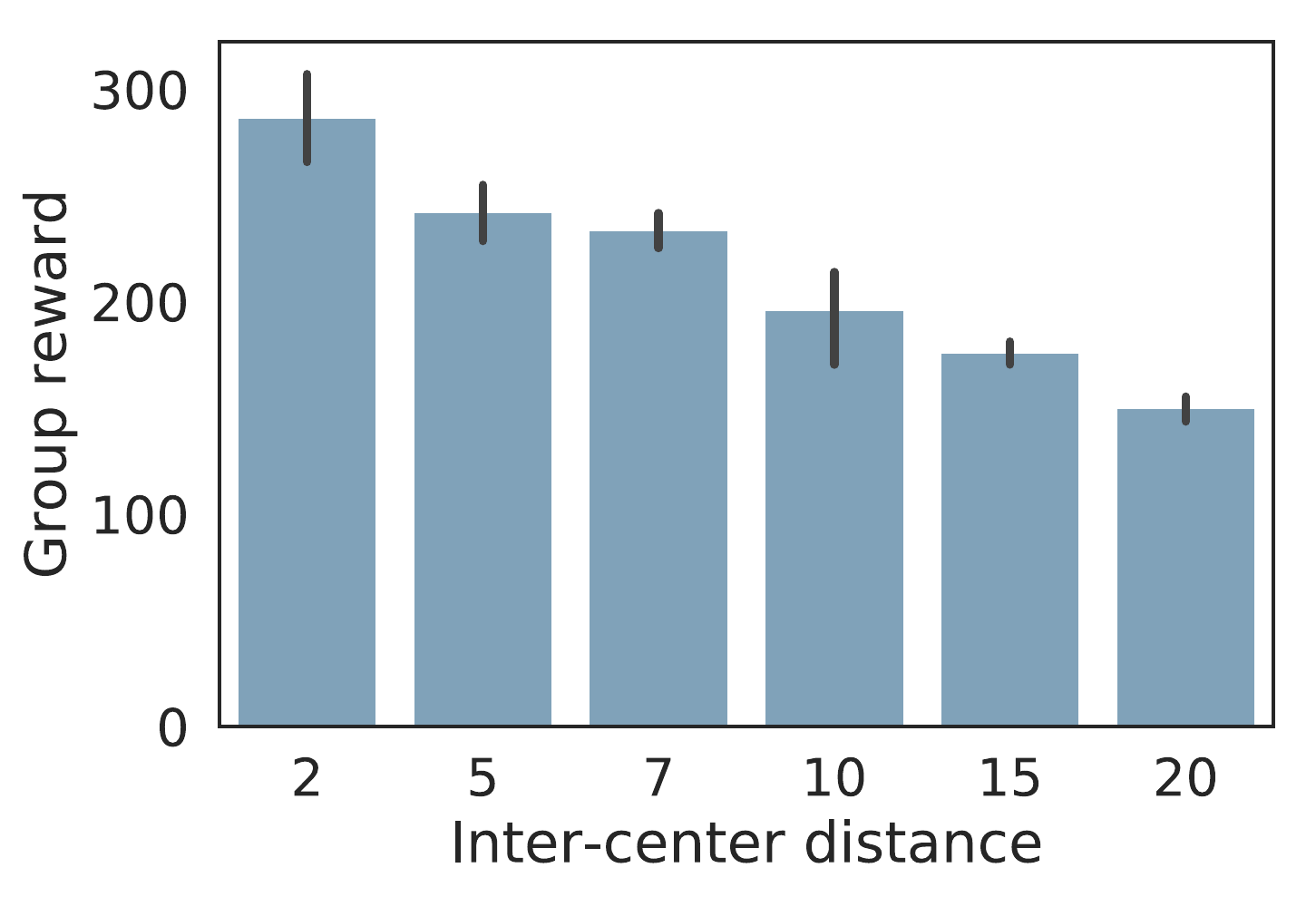}
        \caption{Group reward.}
        \label{fig:geometricchange-return}
    \end{subfigure}
    ~
    \begin{subfigure}{0.32\textwidth}
        \centering
        \includegraphics[width=0.95\linewidth]{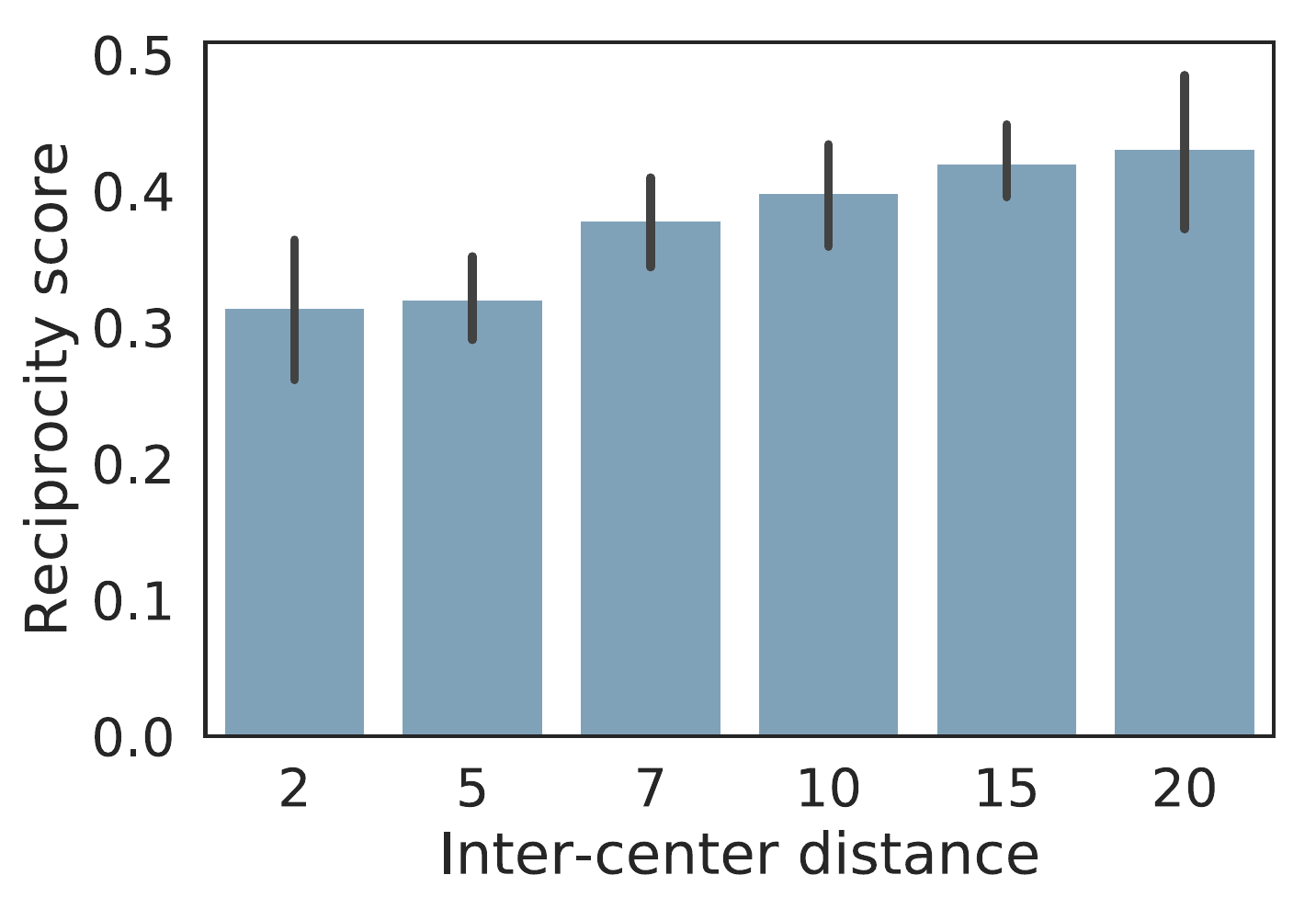}
        \caption{Care reciprocity.}
        \label{fig:geometricchange-recriprocity}
    \end{subfigure} 
    ~
    \begin{subfigure}{0.32\textwidth}
        \centering
        \includegraphics[width=0.95\linewidth]{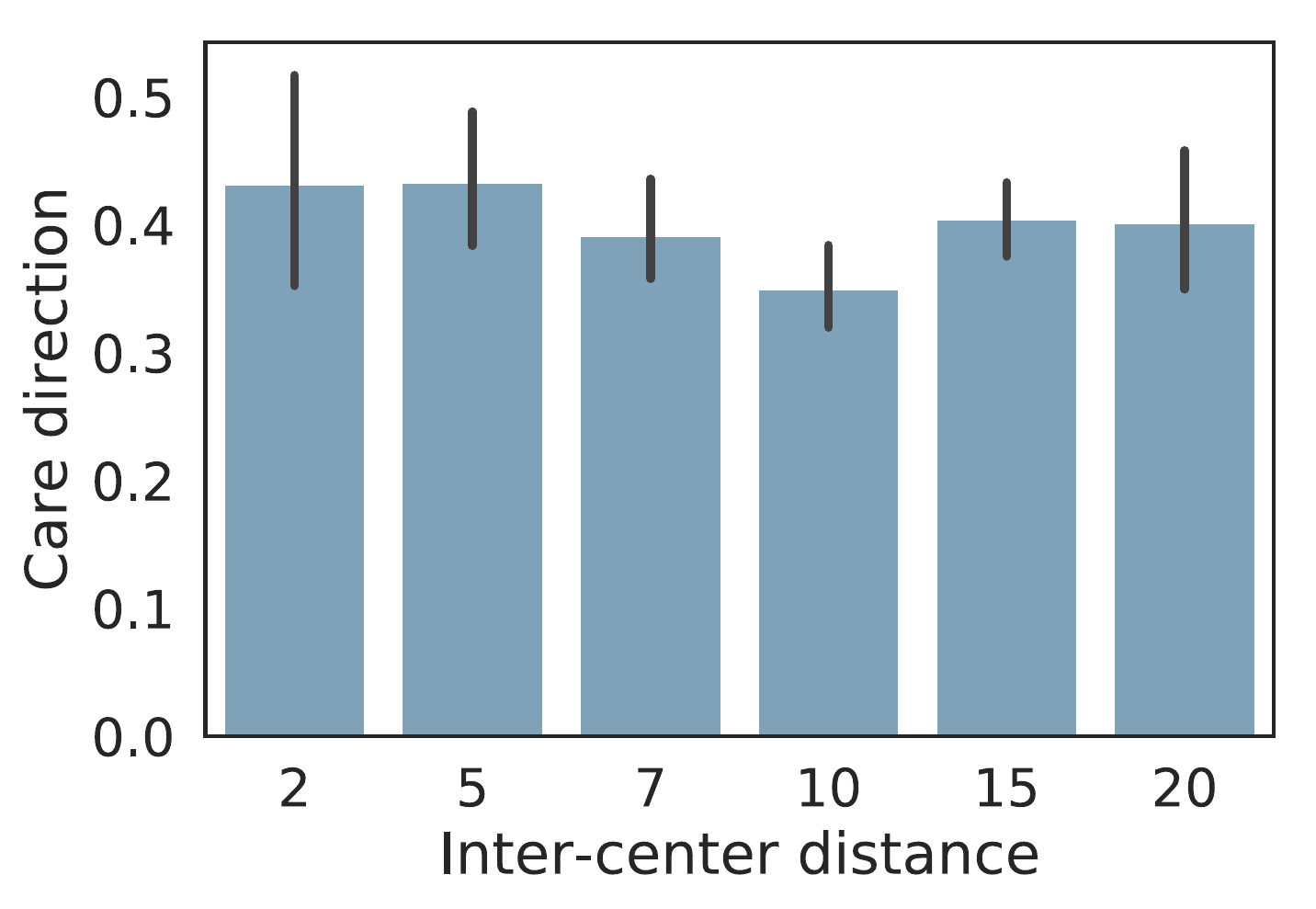}
        \caption{Care direction.}
        \label{fig:geometricchange-direction}
    \end{subfigure}
    \caption{Influence of geometry on social outcome metrics. Increasing the inter-center distance results in lower group reward (a) and higher care reciprocity (b). Metrics are averaged over 8 runs with $95\%$ confidence intervals.}
    \label{fig:geometricchange}
\end{figure*}

\subsection{Maintenance cost}
\label{sec:self_repair}

Our first mechanistic intervention is to enable self-repair. Na\"ively, one might believe that this will cause a uniform uplift in the group reward of the agents, without diminishing the cooperative behavior. Indeed, a supply chain manager, faced with the decision whether to invest in such technology, may well perceive this to be without downsides. However, the dynamics of emergent reciprocity under learning are clearly intricate. As we can see in Figure \ref{fig:evalutate_averagerepairtime}, the introduction of self-repair undermines these dynamics, leading to a significant reduction in care reciprocity, with no significant improvement in group reward. Moreover, we observe that the care that does still occur is predominantly directed upstream rather than the more balanced behavior we observe when self-repair is disabled (Figure \ref{fig:selfrepdirection}). 

This provides a first illustration of the practical benefits of our approach. It is not obvious how to use a network game model to predict the effect of introducing self-repair. By modelling the emergence of care with deep reinforcement learning in a more concrete environment, we can draw nuanced conclusions from grounded interventions.

\subsection{Specialization}\label{sec:spec}

\begin{figure}
        \centering
        \includegraphics[width=0.75\linewidth]{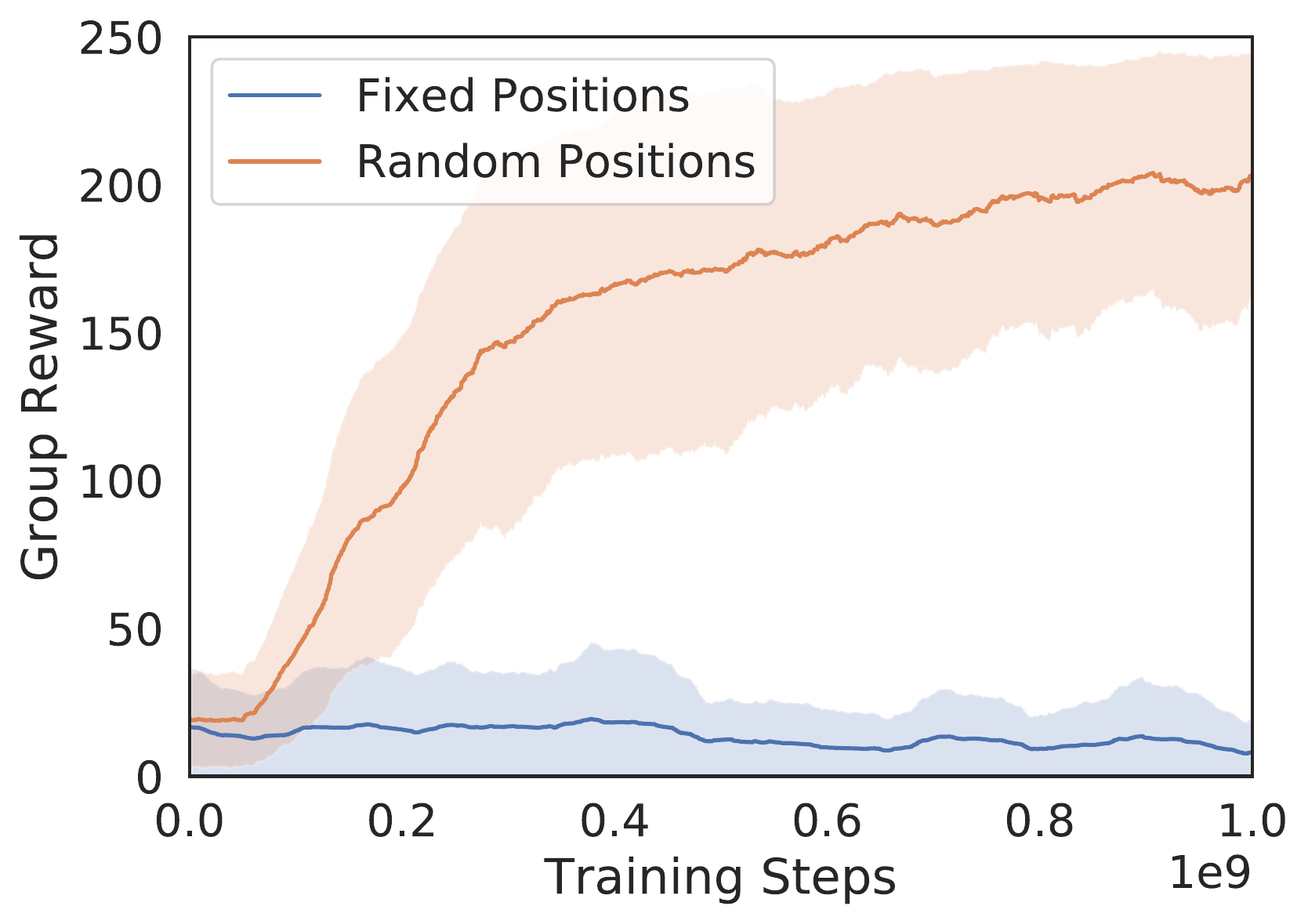}

    \caption{Group rewards during training averaged across $8$ random seeds for agents that are assigned to random processing centers during training versus agents that are assigned to fixed centers.}
    \label{fig:perspectivetaking}
\end{figure}

 \begin{figure*}[ht!]
    \centering
    \begin{subfigure}{0.32\textwidth}
        \centering
        \includegraphics[height=9em]{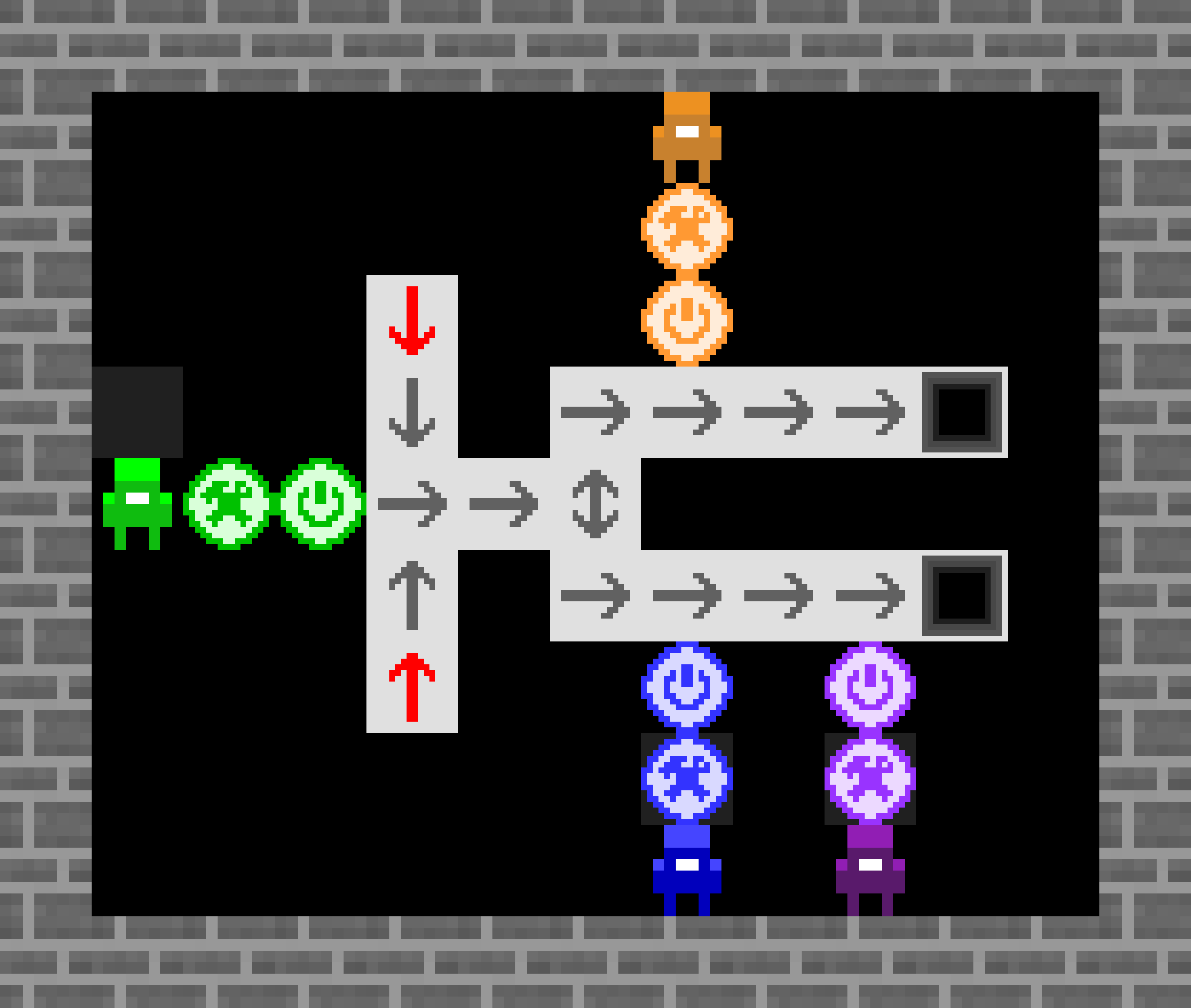}
    \end{subfigure}
    ~
    \begin{subfigure}{0.32\textwidth}
        \centering
        \includegraphics[height=9em]{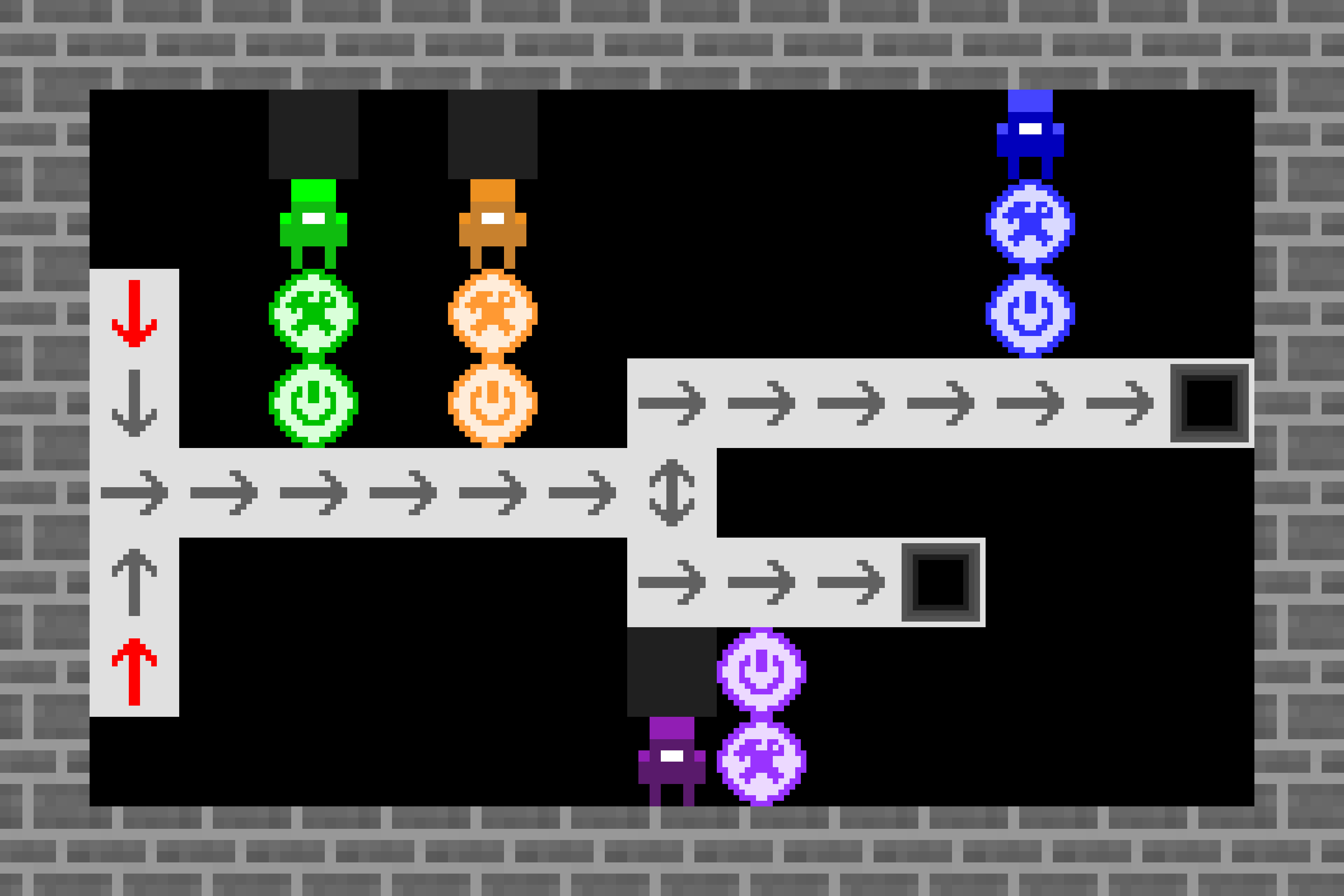}
    \end{subfigure}
    ~
    \begin{subfigure}{0.32\textwidth}
        \centering
        \includegraphics[height=9em]{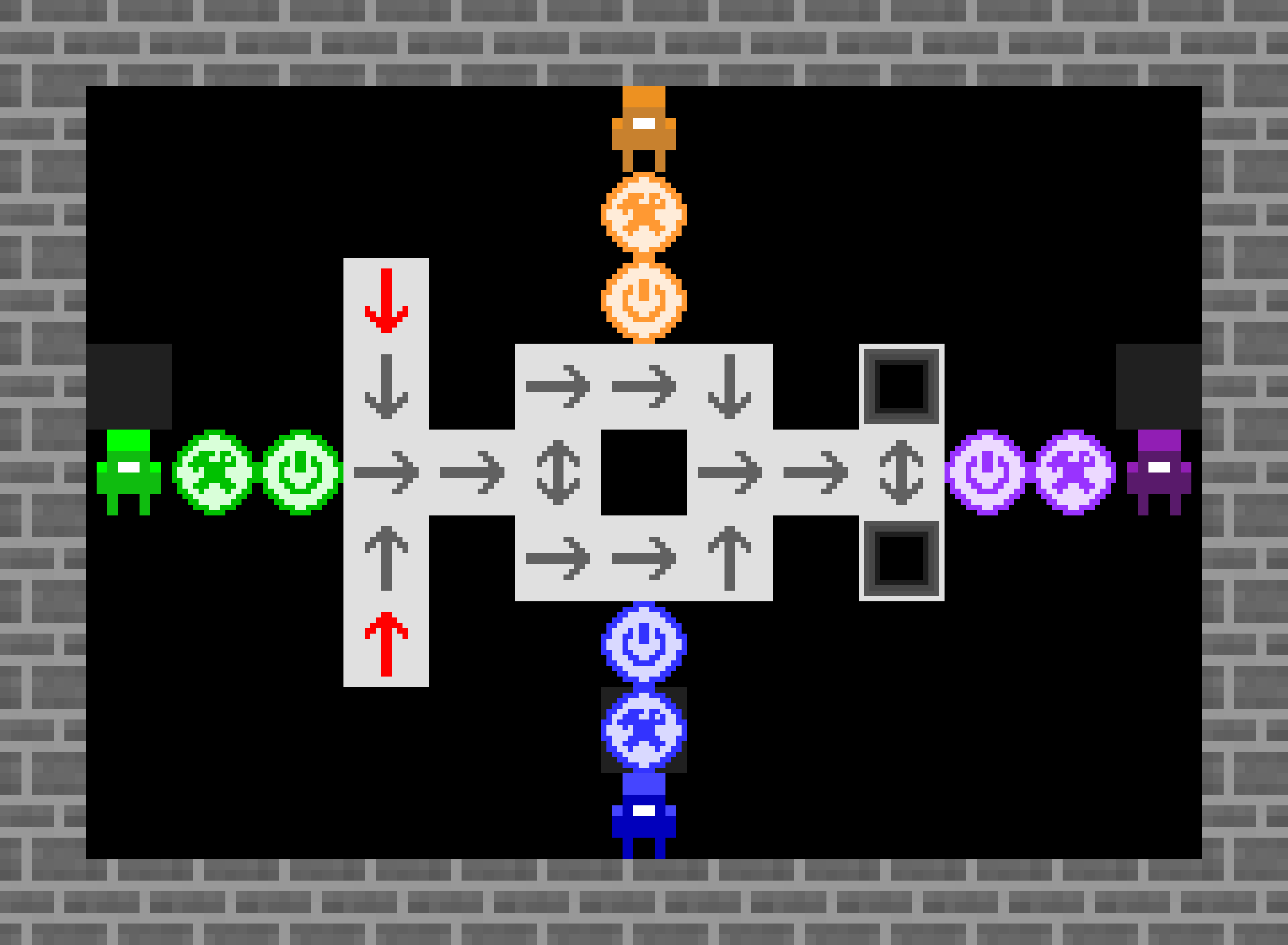}
    \end{subfigure}\\
    \begin{subfigure}{0.32\textwidth}
        \centering
        \includegraphics[width=0.73\linewidth]{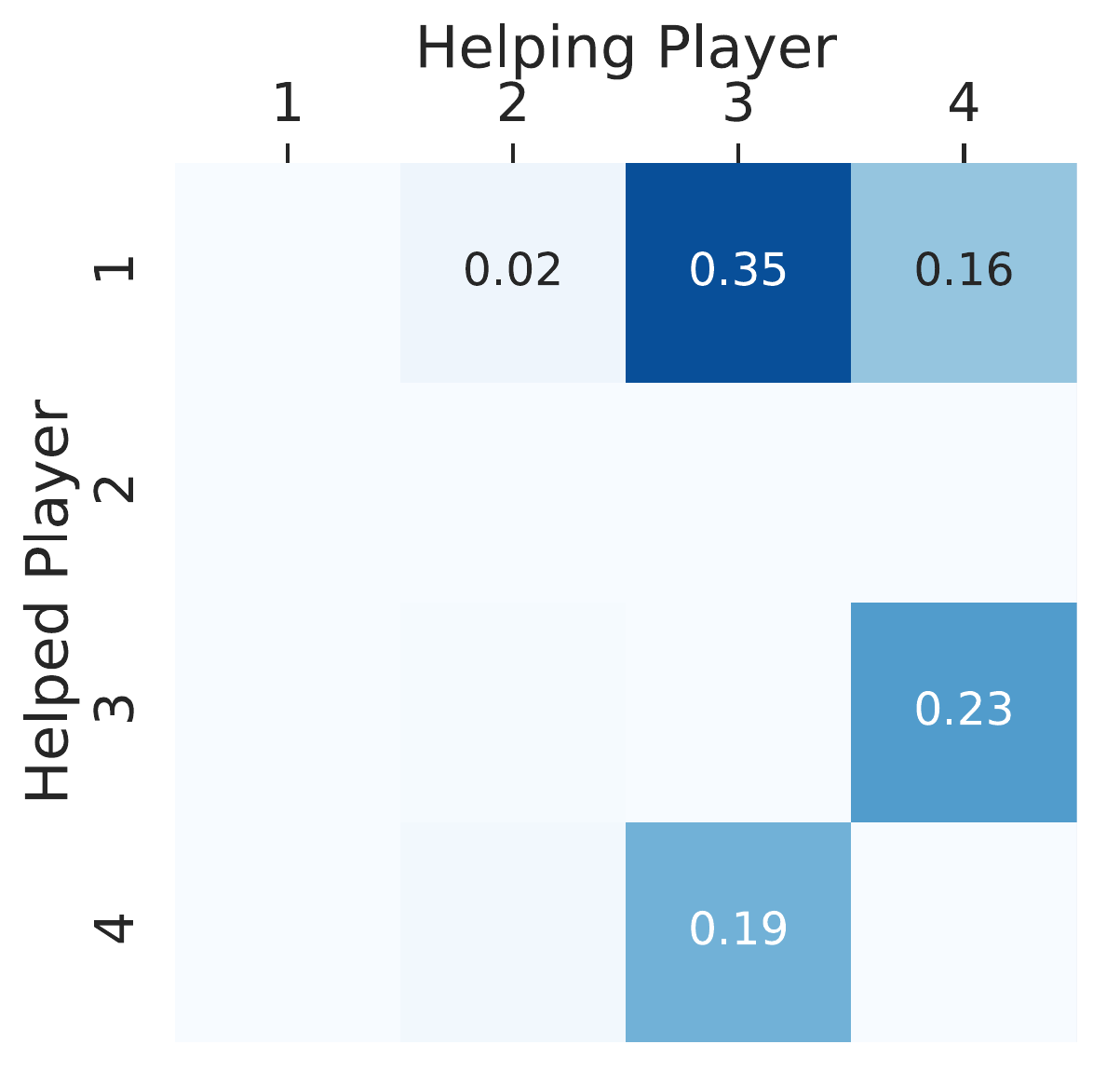}
        \caption{Environment 1.}
        \label{fig:t1}
    \end{subfigure}
    ~
    \begin{subfigure}{0.32\textwidth}
        \centering
        \includegraphics[width=0.73\linewidth]{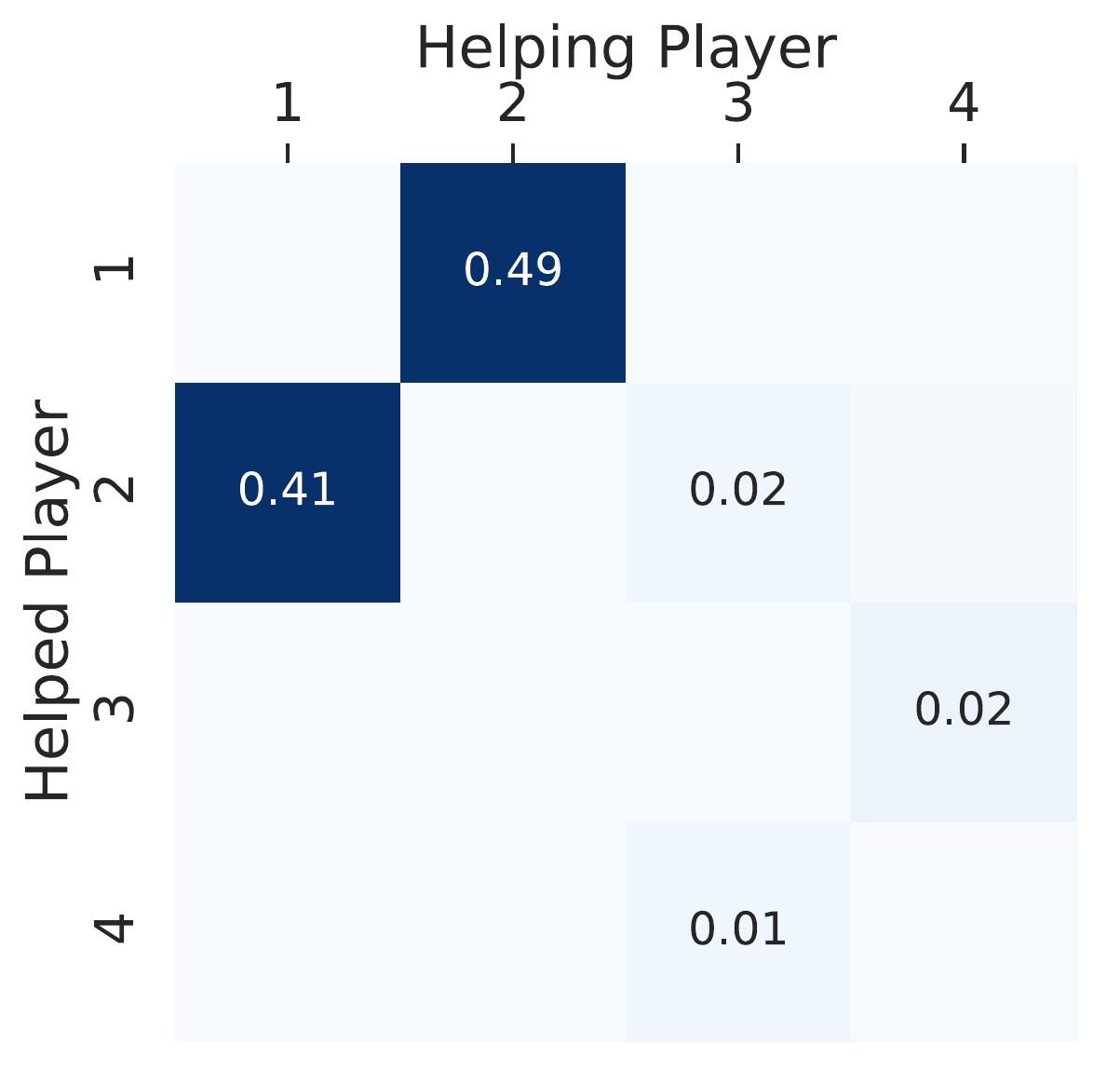}
        \caption{Environment 2.}
        \label{fig:t2}
    \end{subfigure}
    ~
    \begin{subfigure}{0.32\textwidth}
        \centering
        \includegraphics[width=0.73\linewidth]{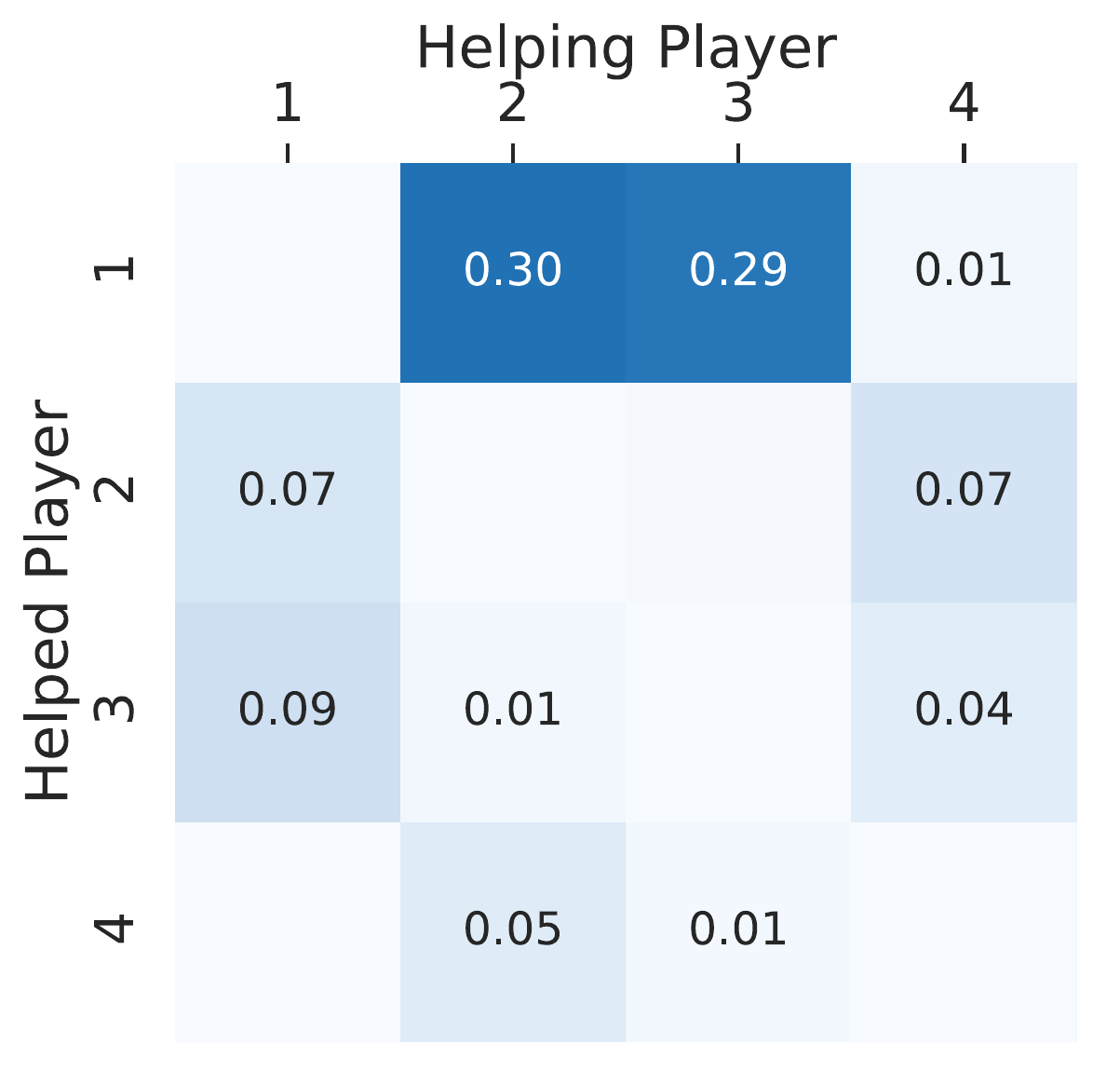}
        \caption{Environment 3.}
        \label{fig:t3}
    \end{subfigure}\\
   
    \caption{Three environments with 4 agents and the corresponding care matrices. The care matrices are averages over 8 different seeds. For improved readability, values below 0.01 are omitted.}
    \label{fig:topologicalchange}
\end{figure*}

\begin{table*}
\centering
\caption{Individual and group rewards, as well as the system efficiency, for the environments visualized in Figure \ref{fig:topologicalchange}. The efficiency corresponds to the percentage of units that leave the environment through a sink node relative to the number of units that enter the environment through a source node.}\label{tab:tjunctions}
\begin{tabular}{@{}l|llll|l|l@{}}
\toprule
Environment    &   $R^1$           & $R^2$           & $R^3$            & $R^4$          & $\sum_i R^i$ & Efficiency  \\ \midrule
Figure \ref{fig:t1} &$125.9 \pm 2.1$ & $4.01 \pm 0.48$ & $55.82 \pm 0.85$ & $48.5 \pm 2.4$ & $234.2 \pm 3.5$ & $26.1\% \pm 2.1 \%$ \\
Figure \ref{fig:t2} & $135.7 \pm 1.3$ & $116.7 \pm 3.3$ & $9.5 \pm 4.8$    & $7.5 \pm 6.2$  & $269.4 \pm 5.1$ & $8.5\% \pm 0.3 \%$ \\
Figure \ref{fig:t3}& $119.4 \pm 5.1$ & $31 \pm 15$     & $31 \pm 14$      & $17 \pm 14$    & $198 \pm 21$ & $8.4\% \pm 6.2\%$\\
\bottomrule
\end{tabular}
\end{table*}

At the start of each episode, the supply chain is populated by four agents. There are several possible strategies for assigning agents from a finite-sized population to each of the four processing centers. So far, we have randomly assigned agents to positions at the start of each episode. Thus the agents must learn policies which generalize across positions. Alternatively, we could encourage specialization by always assigning the same agents to the same positions in the supply chain. This would eliminate the need to learn a single policy that generalizes across positions, and may speed up learning. As such, it is an appealing intervention from the perspective of a mechanism designer. 

However, as we see in Figure \ref{fig:perspectivetaking}, agents only learn to care when they are assigned to different positions during training. In other words, specialization undermines the learning dynamics that beget caring. In effect, specialization leads to overfitting. Agents always appear in the same positions and learn that they can receive some reward by standing on their activation tile. This is a local optimum, for leaving the activation tile has an immediate opportunity cost for processing further units. What's more, there is little benefit in repairing another agent's processing center, unless someone is likely to repair yours: this is the root of the collective action problem. Thus discovering the benefits of care is a hard joint exploration problem. Diversity in role experience helps to solve this problem, at the cost of longer learning times. The link between random role assignment and the selection of just policies has also been studied in political theory, most famously through the ``veil of ignorance'' thought experiment developed by John Rawls~\cite{rawls2009theory}. This connection represents a fruitful area for future research \cite{gabriel2020artificial}.

\subsection{Changing the geometry}\label{sec:geometry}

In contrast to a game-theoretic analysis of the network structure, multi-agent reinforcement learning provides tools for investigating the impact of geometric changes to the environment that are not captured in the abstract graph. 

To demonstrate this, we vary the distance between processing centers and inspect the group reward, care reciprocity and care direction at convergence (Figure \ref{fig:geometricchange}). To have more fine-grained control over the inter-center distance, we use a linear chain rather that the circular chain presented in Figure \ref{fig:environment} (see the Appendix for a visualization of the environment). First, we observe that the group reward decreases as the inter-center distance increases (Figure \ref{fig:geometricchange-return}). 
Interestingly, however, we observe that changing the geometry not only affects the overall reward but also changes the dynamics of care. The longer distance increases the effective ``cost'' of caring which makes the reciprocal nature of care more important (Figure \ref{fig:geometricchange-recriprocity}) while no significant impact is observed on the care direction (Figure \ref{fig:geometricchange-direction}).

In summary, the dynamics of care and the geometry of the environment are intertwined in a non-trivial way. Our model therefore highlights a tension which is invisible in standard network games. Namely, the trade-off between encouraging caring dynamics and increasing group reward. In real-world systems involving a mixture of humans and agents, this trade-off is vitally important: we would not want to simply optimize for group reward at the expense of cooperation \cite{held2006ethics}. Using our approach, one can design mechanisms for the geographical layout of network systems to balance social goods.

\subsection{Changing the topology}\label{sec:topology}
Thus far, we have analyzed the social outcomes in supply chains where the underlying abstract graph structure is linear. In practice, networks often have more complex topologies. To investigate the effect of topology on social outcomes, we introduce three environments with the same number of agents but different underlying graphs that govern how units flow in the supply chain, see Figure \ref{fig:topologicalchange}.  

In Table \ref{tab:tjunctions}, we show the individual and cumulative rewards in each environment. In all three environments, agent 1 accumulates most reward on average because all other agents depend on this processing center for reward. Rewards for the other agents depend strongly on the topology. In the first environment (Figure \ref{fig:t1}), the supply chain branches out after center $1$ with each branch randomly receiving half of the units. In the top branch, agent 2 earns relatively little reward as no other agent has a strong incentive to care for its processing center. In the bottom branch, however, agents 3 and 4 both earn close to half of the reward of the first agent as they each have an incentive to care for each other. Agent 4 does so because agent 3 is upstream, while agent 3 has an incentive to reciprocate that care to ensure that agent 4 keeps repairing. 

The importance of reciprocity is also observed in the social outcomes of the second supply chain (Figure  \ref{fig:t2}). Here the chain branches out after two agents, and those two agents thus depend on each other for care. By contrast, the last two agents are in different branches of the tree, so are not interdependent and thus fail to form a reciprocal pair. 

Finally, in the third supply chain (Figure \ref{fig:t3}), the chain branches out after the first agent but reconnects again before the last agent. Here learning is more unstable. Agents 2, 3 and 4 all earn some reward but the amount of collected reward varies strongly between different runs (see Table \ref{tab:tjunctions}). This is because there are multiple stable social outcomes that the system randomly converges to (see the Appendix for care matrices of the individual runs). Averaged across all 8 runs, we observe the highest care from agent 2 and 3 to agent 1, intuitively sensible since they both depend directly on agent 1 for reward. In turn, we observe that some of this help is reciprocated by agent 1. Finally, we see that though the last agent is able to establish some reciprocal care with agents 2 and 3, it is receiving the least reward.

A mechanism designer may be interested in creating a supply chain that maximizes the number of units that are successfully processed by the system as a whole rather than maximizing the sum of individual rewards. Hence, we compute the efficiency of the system (see Table\,\ref{tab:tjunctions}), defined as the number of units that leave the system at any of the sink nodes relative to the the number of units that enter the system.
Interestingly, in these complex topologies, the system efficiency does not precisely correlate with group reward. Indeed, we observe the best overall system efficiency in environment 1. Though environment 2 yields the highest group reward, most units are discarded before they reach the sink tiles, resulting in a low number of units processed by the system overall. This highlights the value of conducting detailed simulations admitting multiple metrics.

\section{Discussion}

Games with similar topological structures have been considered in the cooperative game theory literature~\cite{markus2012shapley,littlechild1973simple}. Cost-tree games, also called irrigation games, are a class of transferable utility network games where each edge of a directed tree has an associated cost (e.g. for maintenance) that should be shared across the users (e.g. farmers in an irrigation system). The Shapley solution \cite{shapley1951notes} predicts that the costs corresponding to each edge are shared equally by all users that depend on that edge. This is at odds with the care reciprocity we observe in our experiments. In our spatio-temporally complex setting, the assumption of transferable utility is too reductive. Our method allows to extend beyond such assumptions by way of empirical game theory \cite{wellman2006methods}, reaching new conclusions. 

There are several natural avenues for extending our contribution. Firstly, it would be valuable to extend the dynamics of our environment to incorporate more detail of supply chains, along the lines of the Beer game. Secondly, it would be fruitful to explore the consequences of our methods in other complex environments with graph structure, such as $2$- or $3$- dimensional realizations of an irrigation system. This may pave the way for a wider class of metrics, and allow us to investigate a broader range of mechanisms that promote cooperative behavior. Along the same lines, it would be useful to extend our analysis to larger numbers of agents in interaction, where one might be able to observe phase transition-like effects under learning more clearly. 

More broadly, our work raises several questions that may spur future investigation in this field. How can we automate the design of mechanisms that depend on structural changes to the world? This question intersects with the lively field of procedural generation for reinforcement learning environments~\cite{DBLP:journals/corr/abs-1806-10729, DBLP:journals/corr/abs-1807-01281, 10.5555/3306127.3331971}. It also relates to recent work on learning to teach~\cite{DBLP:journals/corr/abs-1810-00147, DBLP:journals/corr/FachantidisTV17, DBLP:journals/corr/abs-1805-07830} and learning to incentivize~\cite{zheng2020ai, lio2020, lupu2020gifting}. A complementary direction would involve validating our model by comparison with experiments involving human participants: to what extent are the conclusions we draw about mechanistic interventions borne out in the behavior of humans? Tackling such questions in progressively greater detail may yield important insights into the management of real-world collective action problems in the years to come. 

\section*{Acknowledgements} We would like to thank Theophane Weber, Kevin McKee and many other colleagues at DeepMind for useful discussions and feedback on this work.

\bibliographystyle{ACM-Reference-Format} 
\bibliography{supplychainbib}

\clearpage
\appendix
\setcounter{figure}{0}
\renewcommand{\thefigure}{A\arabic{figure}}

\begin{figure*}[h]
    \begin{subfigure}{0.2\textwidth}
        \centering
        \includegraphics[height=4.5em]{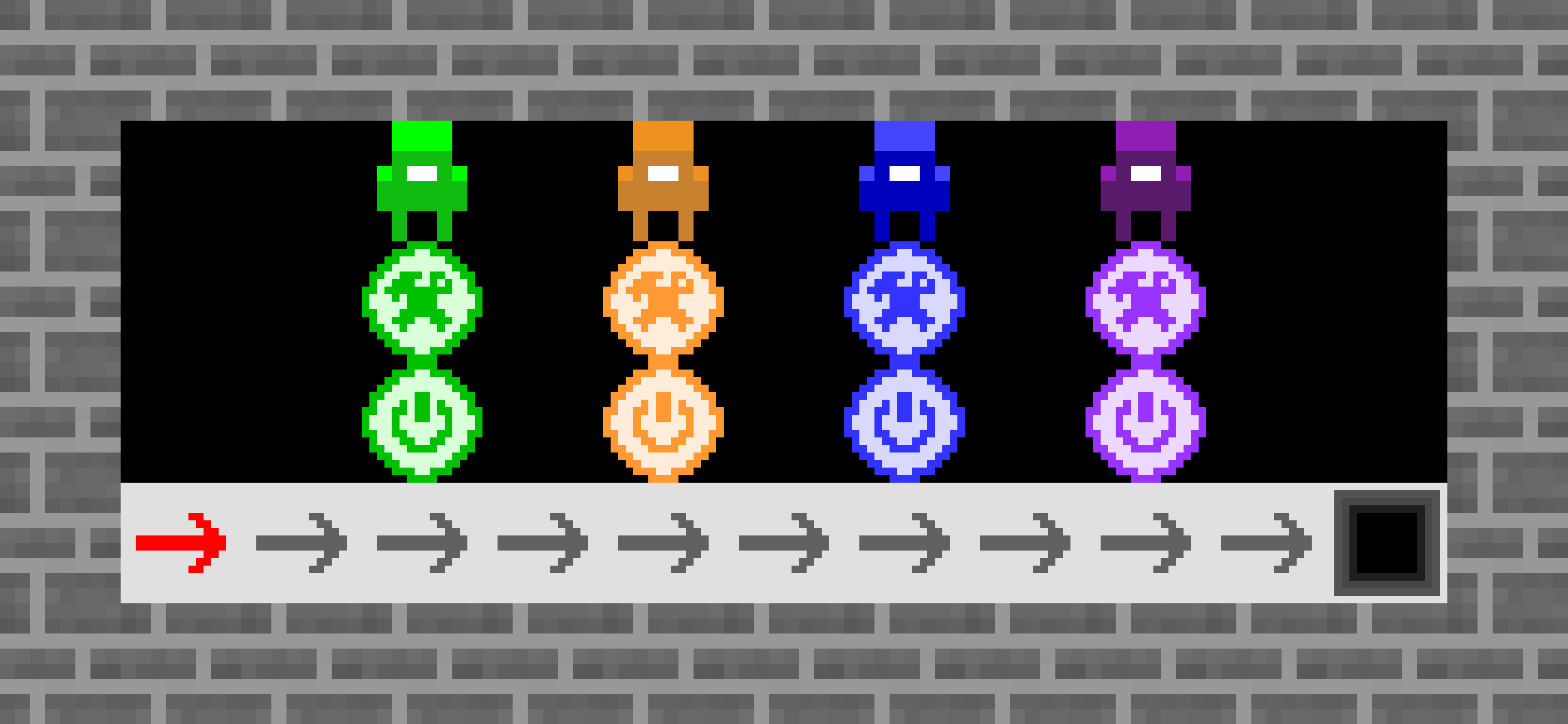}
        \caption{d = 2.}
    \end{subfigure}
    \hfill
    \begin{subfigure}{0.3\textwidth}
        \centering
        \includegraphics[height=4.5em]{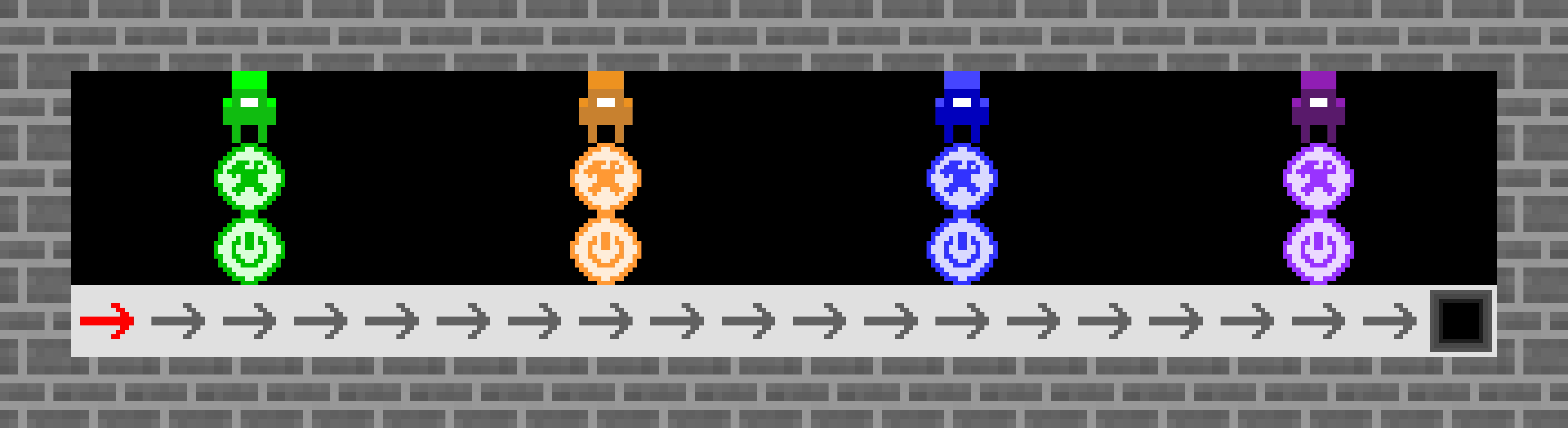}
        \caption{d = 5.}
    \end{subfigure}
    \hfill
    \begin{subfigure}{0.40\textwidth}
        \centering
        \includegraphics[height=4.5em]{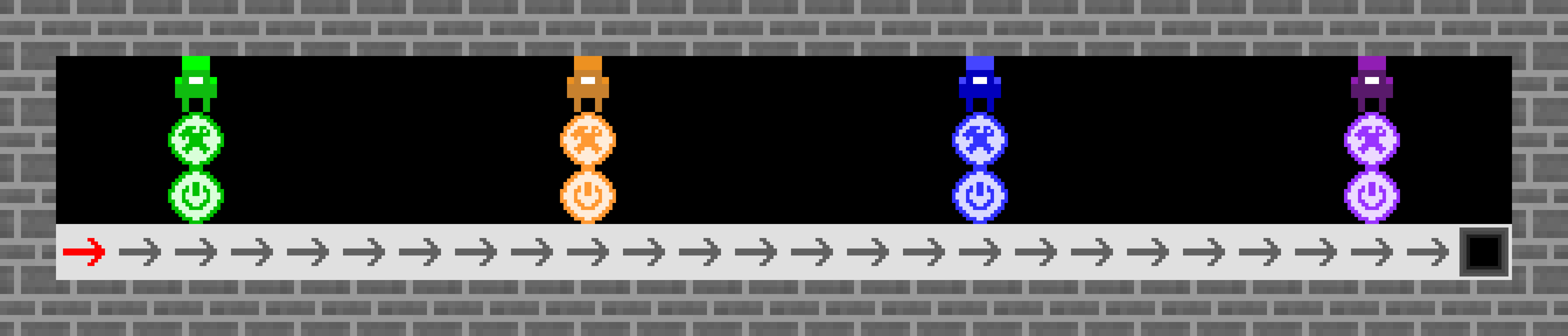}
        \caption{d = 7.}
    \end{subfigure}
    \caption{Examples of the linear layout at varying inter-center distances.}
    \label{fig:linear_environment}
\end{figure*}

\section{Environment}

\subsection{Linear supply chains}
In Section \ref{sec:geometry} in the main paper, where we vary the inter-center distance, we use a linear version of the supply chain instead of the canonical circular one shown in Figure \ref{fig:environment}. Examples of this type of environment for three different inter-center distances can be found in Figure \ref{fig:linear_environment}. Note only the geometry differs between the linear and circular environments while the topology remains the same (a chain with four nodes).

\begin{figure*}[h]
    \begin{subfigure}{0.32\textwidth}
        \centering
        \includegraphics[width=\linewidth]{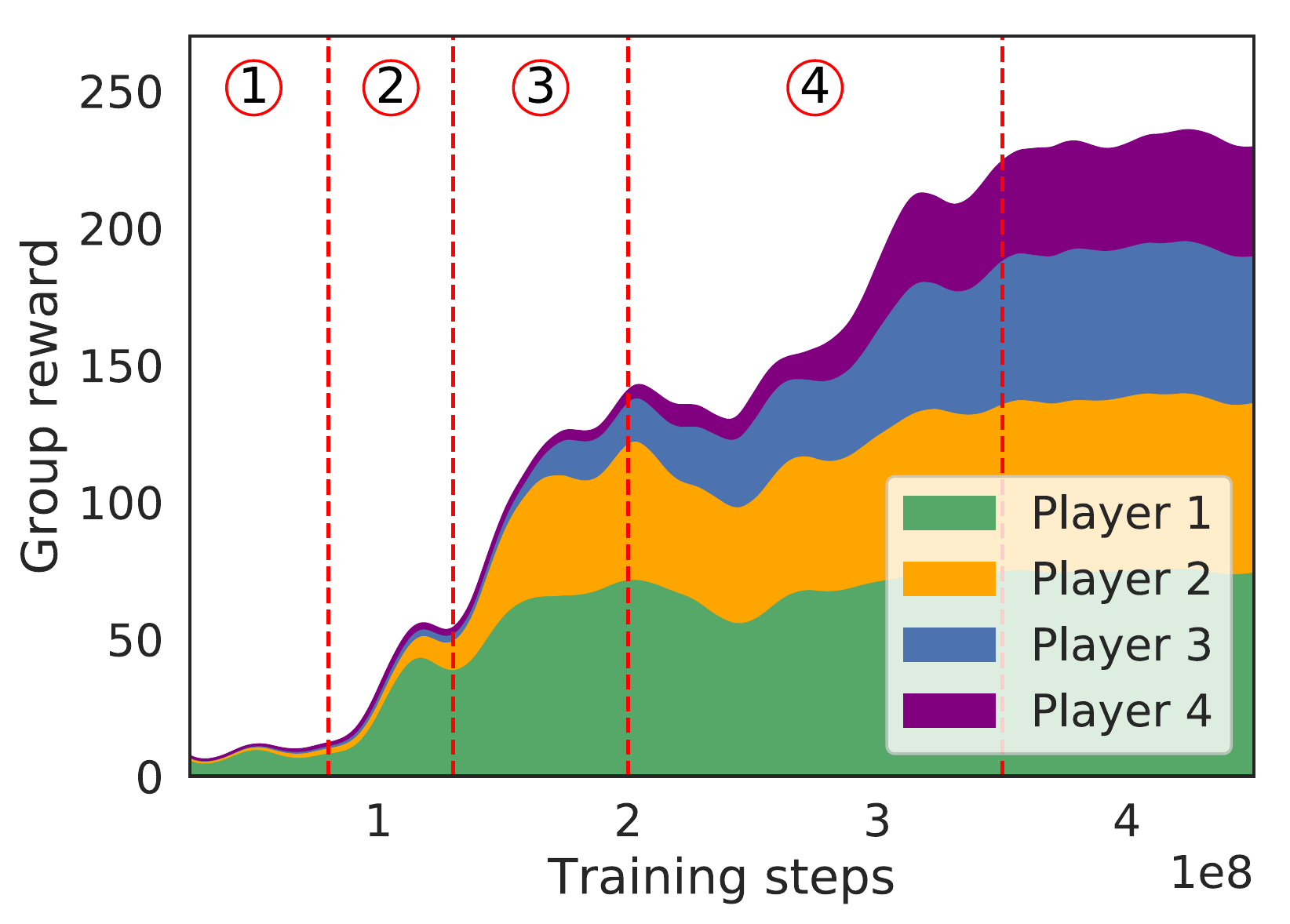}
        \caption{Individual reward per agent stacked for group reward.}
    \end{subfigure}
    \;
    \begin{subfigure}{0.32\textwidth}
        \centering
        \includegraphics[width=\linewidth]{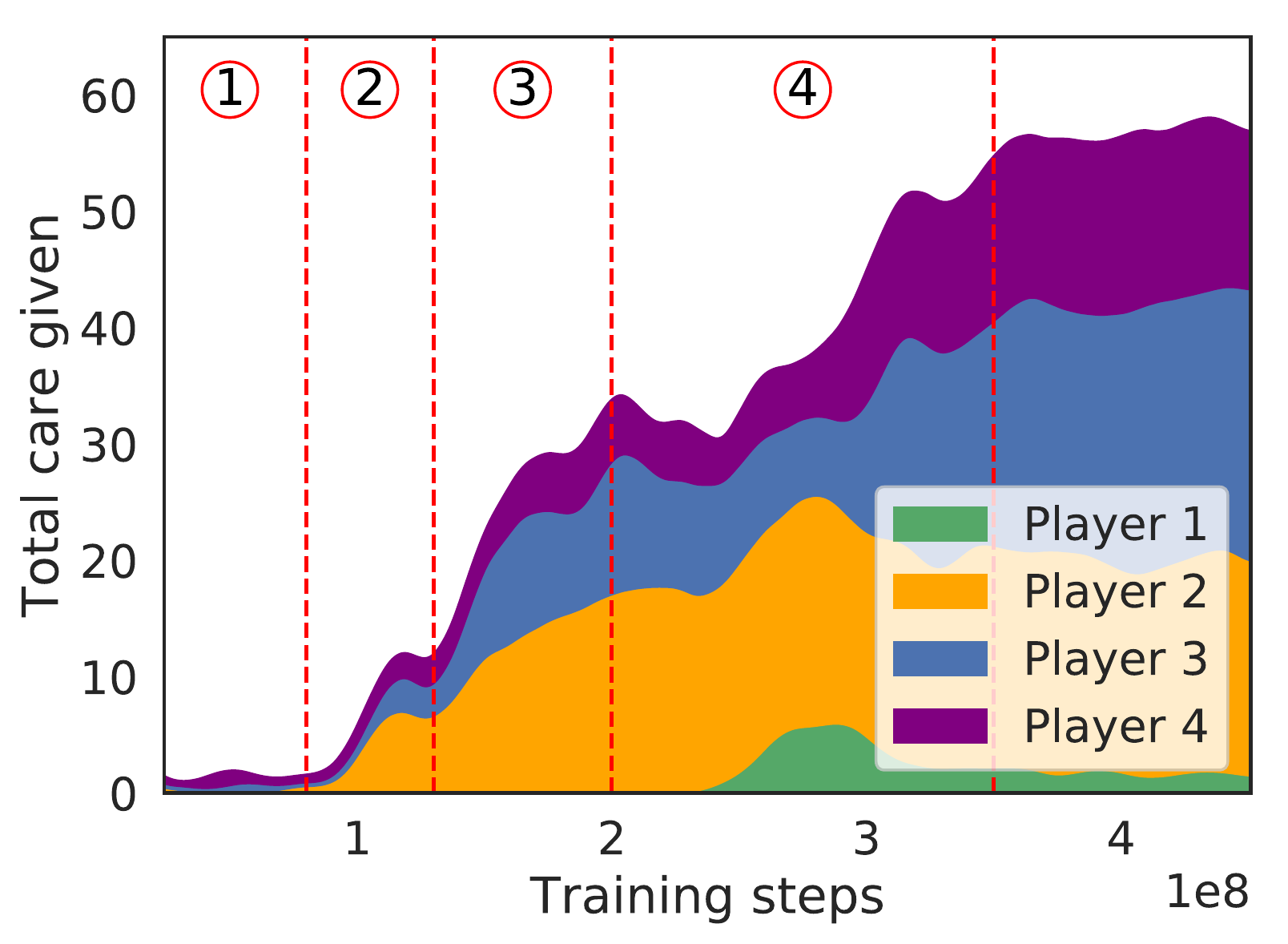}
        \caption{Care given per agent stacked for total care given.}
    \end{subfigure}
    \;
    \begin{subfigure}{0.32\textwidth}
        \centering
        \includegraphics[width=\linewidth]{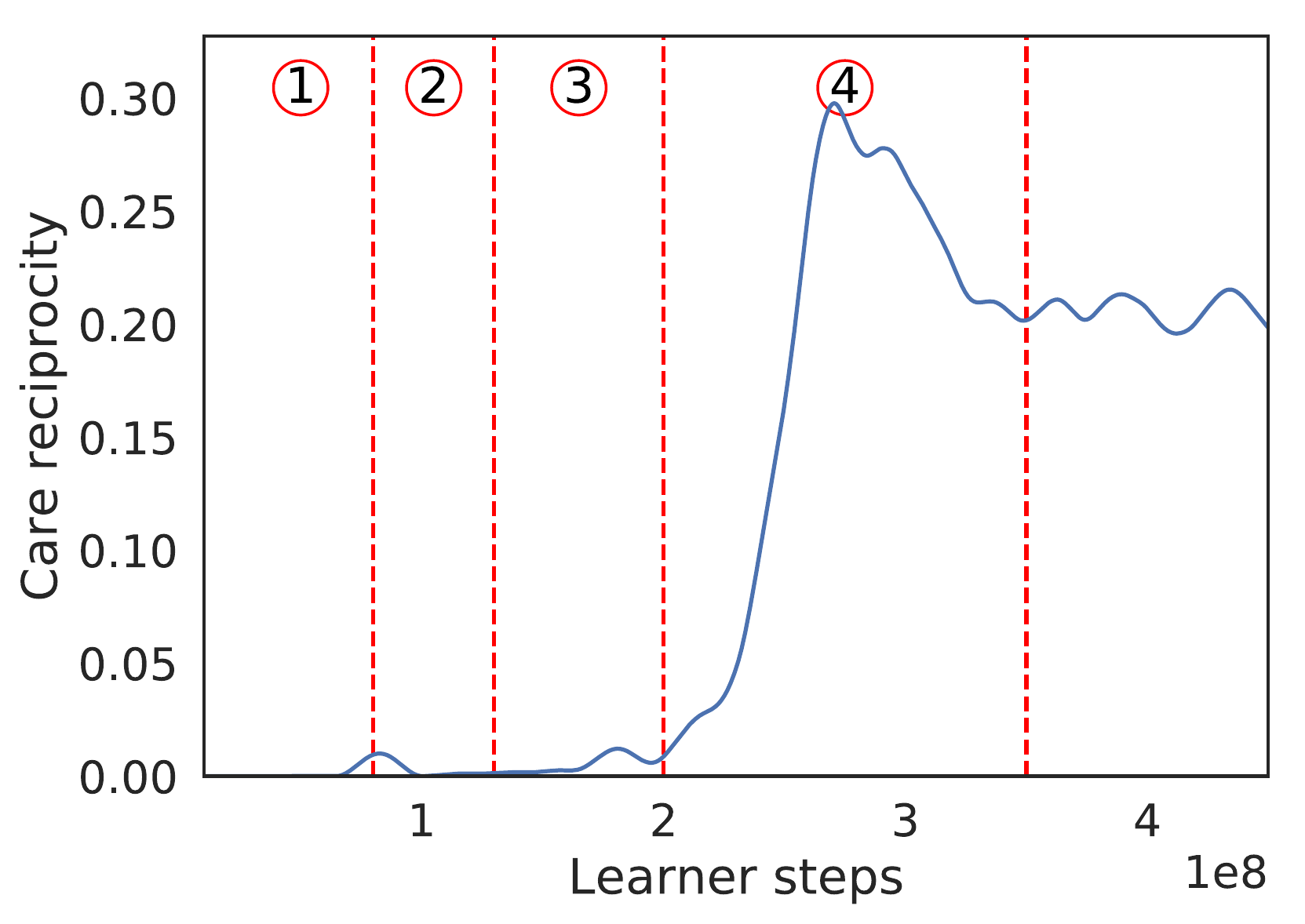}
        \caption{Care reciprocity.}
    \end{subfigure}
    \caption{Evolution of social outcome metrics over the course of training on a circular chain with four agents and self-repair disabled. Please see Figure  \ref{fig:learning_curves_carematrices_2} for the care matrices at the end of each phase.}
    \label{fig:learning_curves_reciprocity_2}
\end{figure*}

\begin{figure*}[h]
    \begin{subfigure}{0.24\textwidth}
        \centering
        \includegraphics[width=\linewidth]{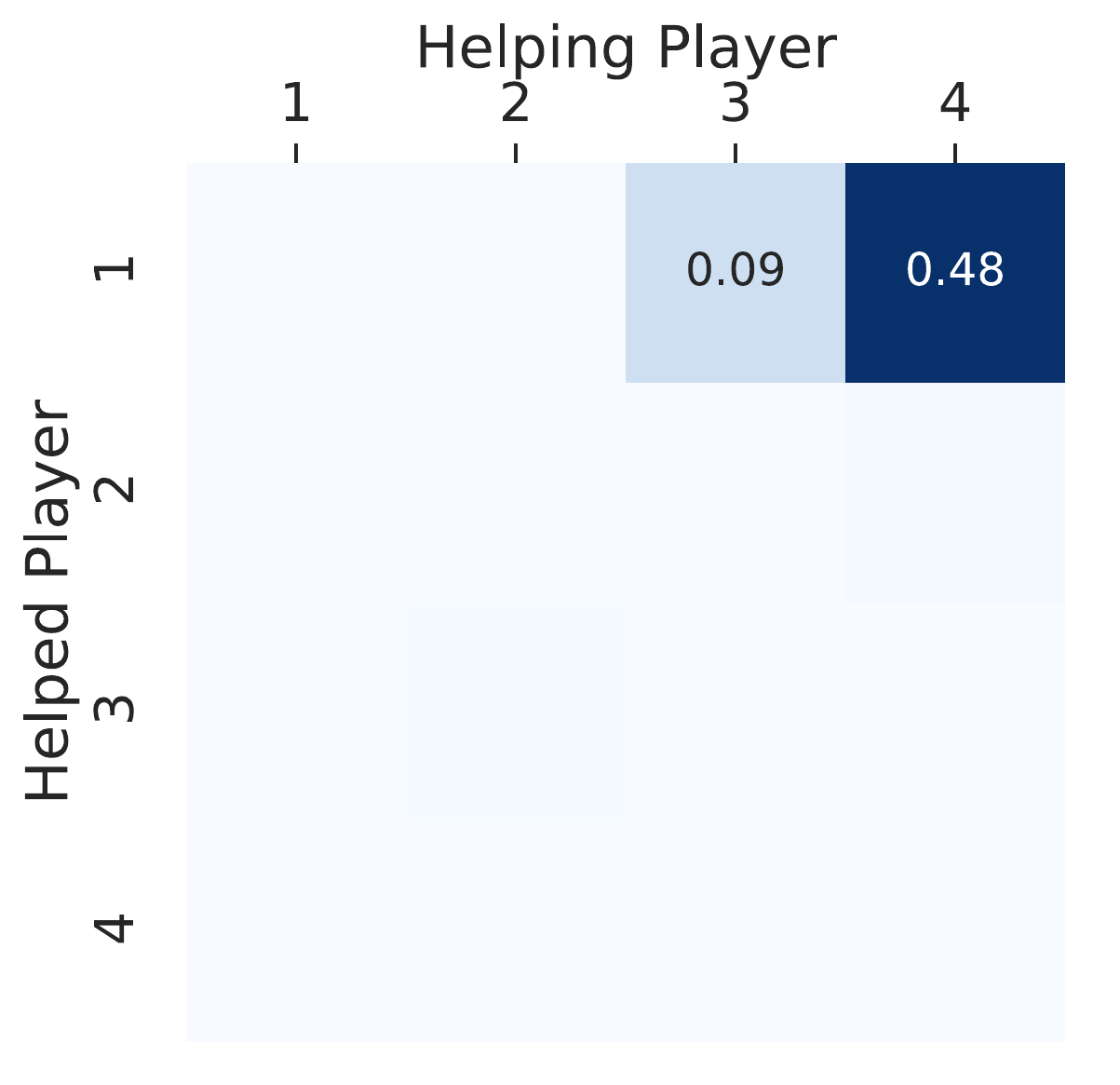}
        \caption{End of phase 1.}
    \end{subfigure}\;
    \begin{subfigure}{0.24\textwidth}
        \centering
        \includegraphics[width=\linewidth]{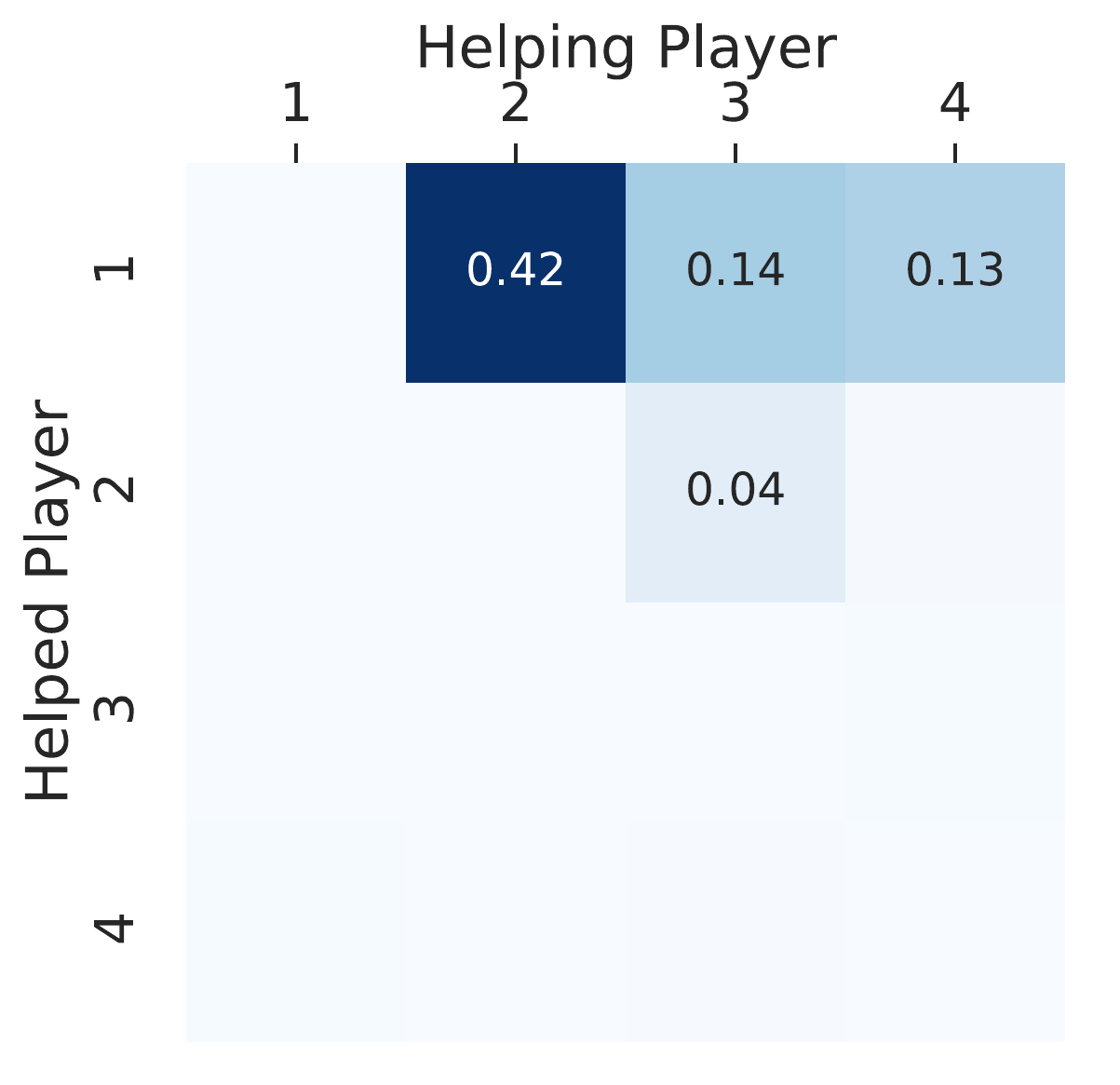}
        \caption{End of phase 2.}
    \end{subfigure}\;
    \begin{subfigure}{0.24\textwidth}
        \centering
        \includegraphics[width=\linewidth]{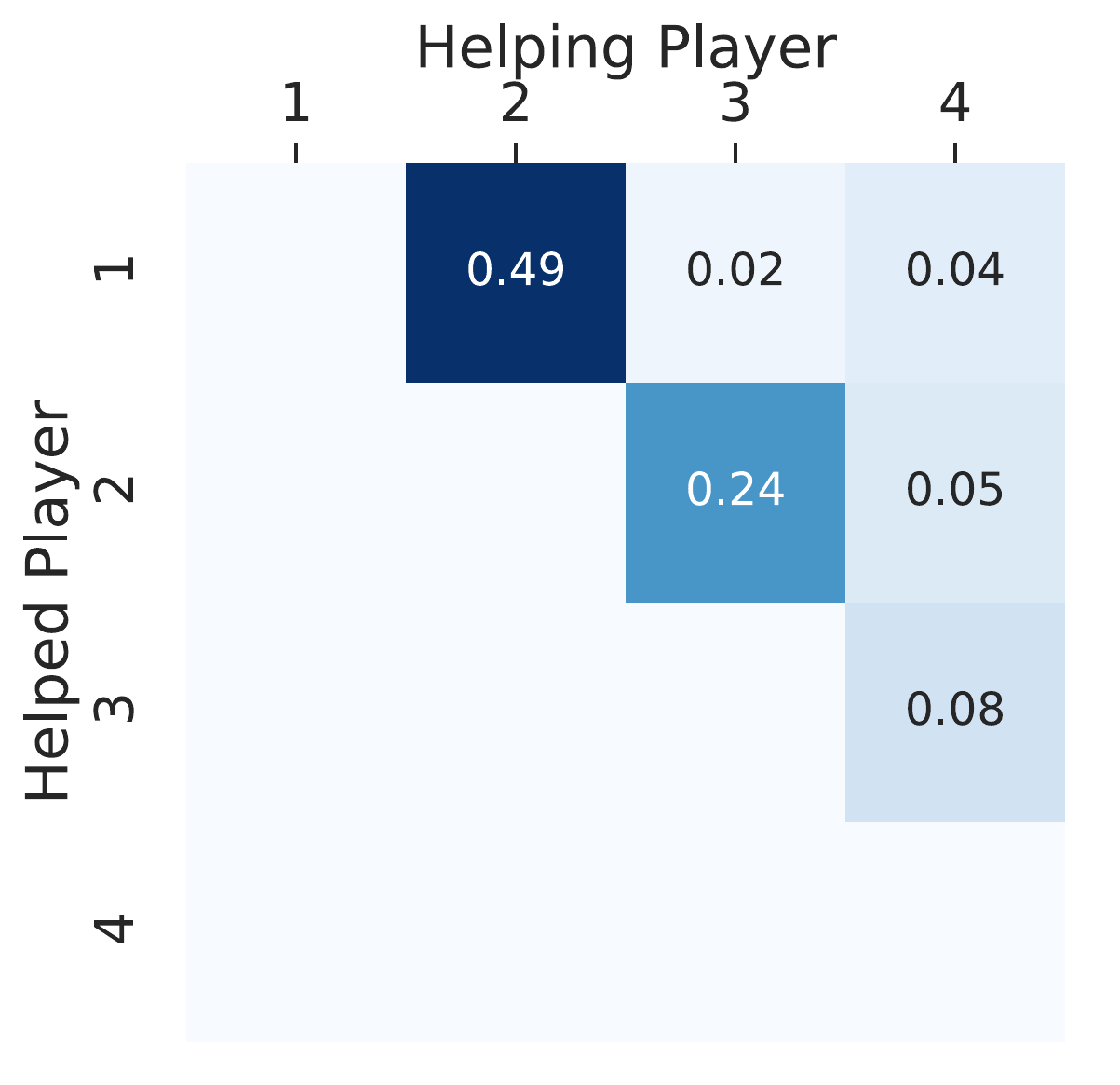}
    \caption{End of phase 3.}
    \end{subfigure}\;
    \begin{subfigure}{0.24\textwidth}
        \centering
        \includegraphics[width=\linewidth]{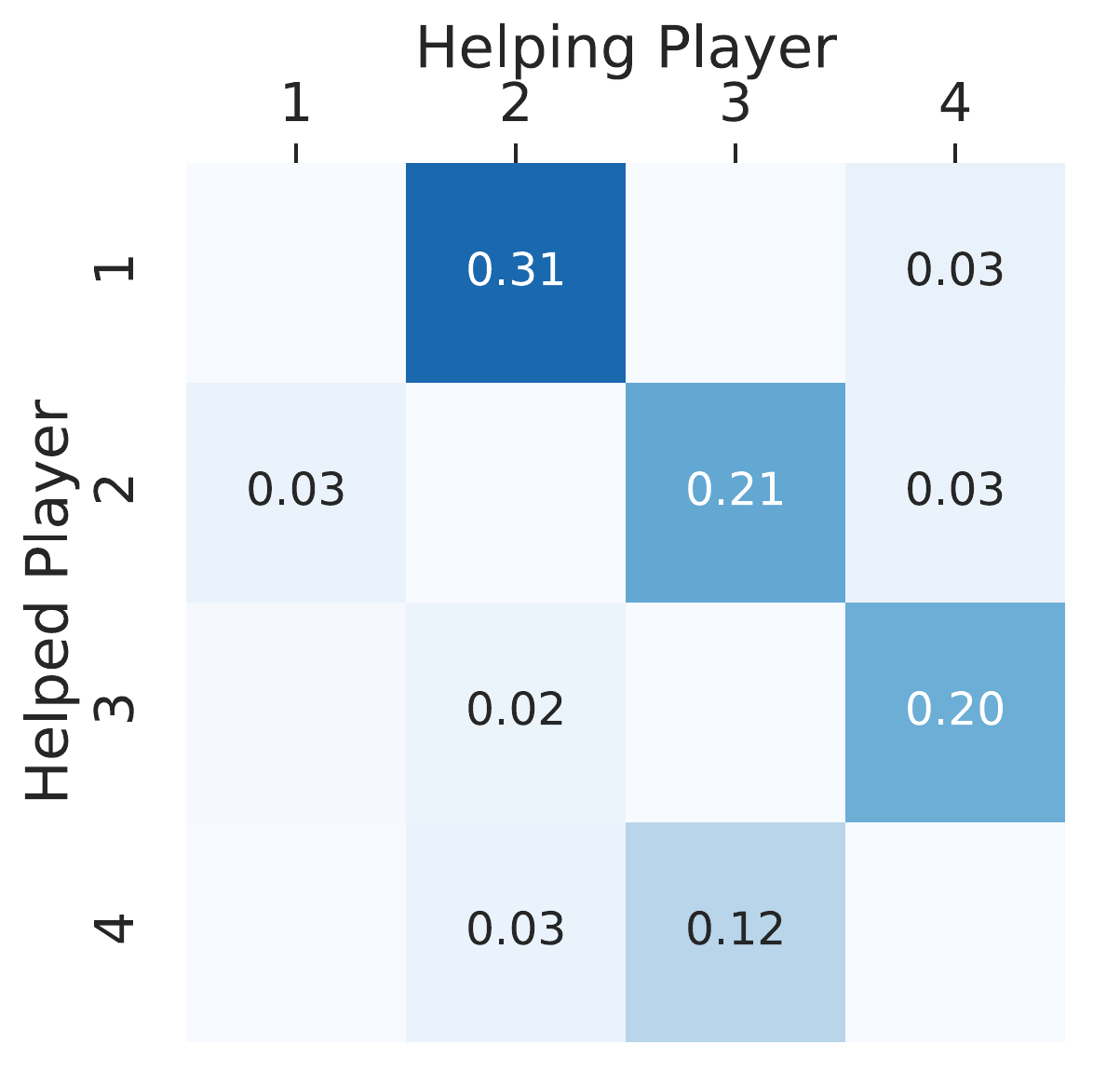}
        \caption{End of phase 4.}\label{fig:baseline_2}
    \end{subfigure}
    \caption{Care matrices at the end of the four distinct learning phases corresponding to Figure \ref{fig:learning_curves_reciprocity_2}. For improved readability, values below 0.01 are omitted.}
    \label{fig:learning_curves_carematrices_2}
\end{figure*}

\section{Additional Results}

\subsection{Emergence of care}
In Section \ref{sec:emergence} in the main text, we discuss the learning dynamics and the emergence of care for a single training run. However, due to the abrupt nature of learning, there are large difference between learning dynamics across individual runs with different random seeds. In Figure \ref{fig:learning_curves_reciprocity_2}, we observe the learning curves for a second typical run while the care matrices at the end of each learning phase can be found in Figure \ref{fig:learning_curves_carematrices_2}. In phase $1$ (ending after $\approx 0.8\cdot 10^8$), we observe that, in contrast to the run highlighted in Section \ref{sec:emergence}, it is now the last agent that first learns to care for the first agent. In the second phase (ending after $\approx 1.3\cdot 10^8$) this behavior is also adopted by the second and the third agent, resulting in a similar care matrix to what we observed at the end of the second phase in the other example run. A large difference between both runs, however, emerges during the third phase (ending after $\approx 2\cdot 10^8$). While we previously observed a large increase in reciprocity during this phase because the first agent learned to reciprocate the care given by the second agent, we now observe that the second agent is instead receiving care from the third agent (and to a lesser extend the third agent from the last agent). Consistent reciprocal care in this run only emerges during the last phase (ending after $\approx 3.5\cdot 10^8$) when the third agent learns to reciprocate the care of the fourth agent. From the perspective of a mechanism designer, it is important to understand which different stable outcomes exist and, if necessary, implement geometric, topological or functional changes to the environment to ensure that the system converges to a desired outcome. 

\subsection{Care matrices of individual runs for one T-junction supply chain}
\begin{figure*}[h]
    \begin{subfigure}{0.23\textwidth}
        \centering
        \includegraphics[width=\linewidth]{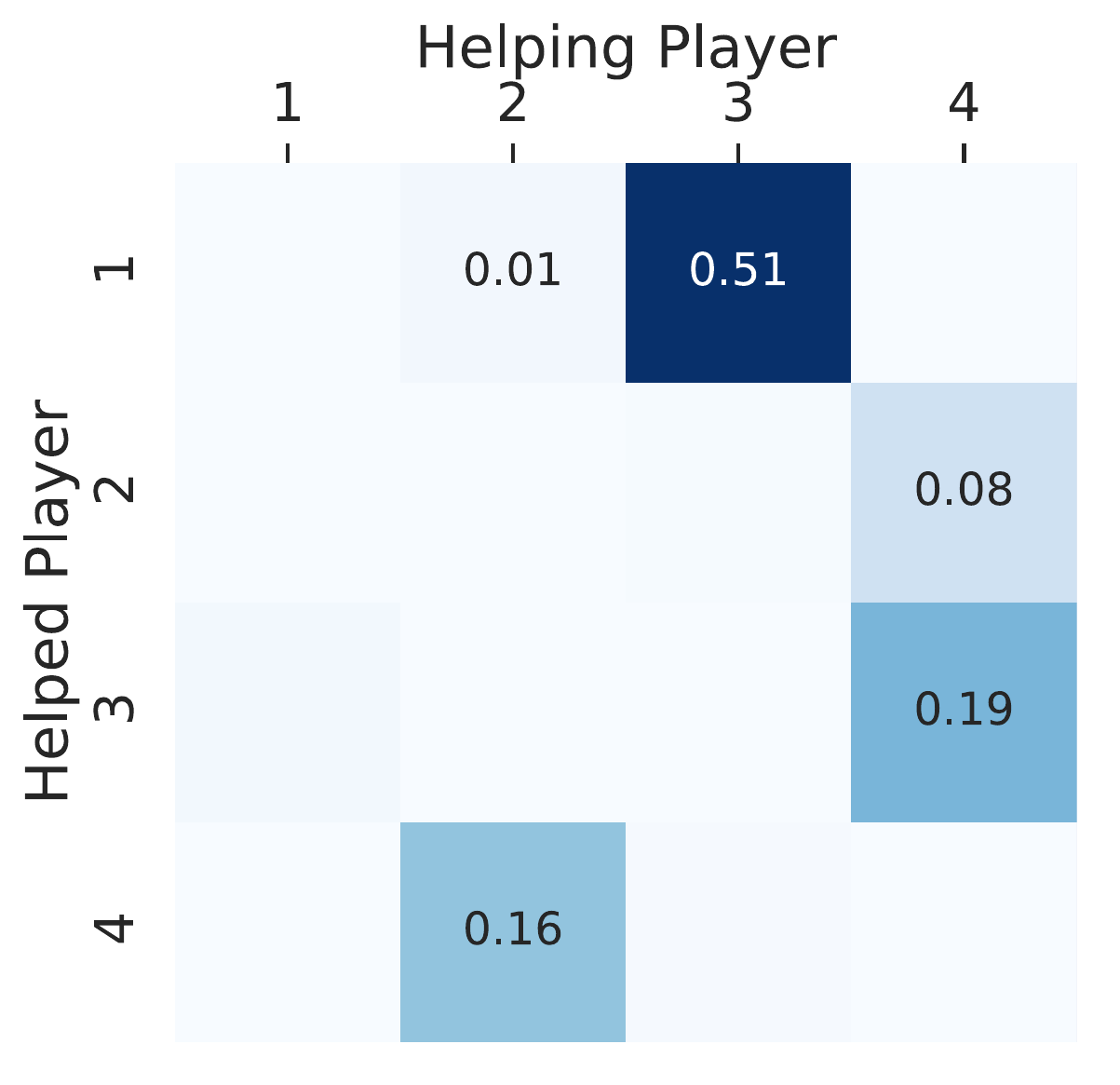}
        \caption{Care matrix run 1.}\label{fig:teqrun1}
    \end{subfigure}
    \;
    \begin{subfigure}{0.23\textwidth}
        \centering
        \includegraphics[width=\linewidth]{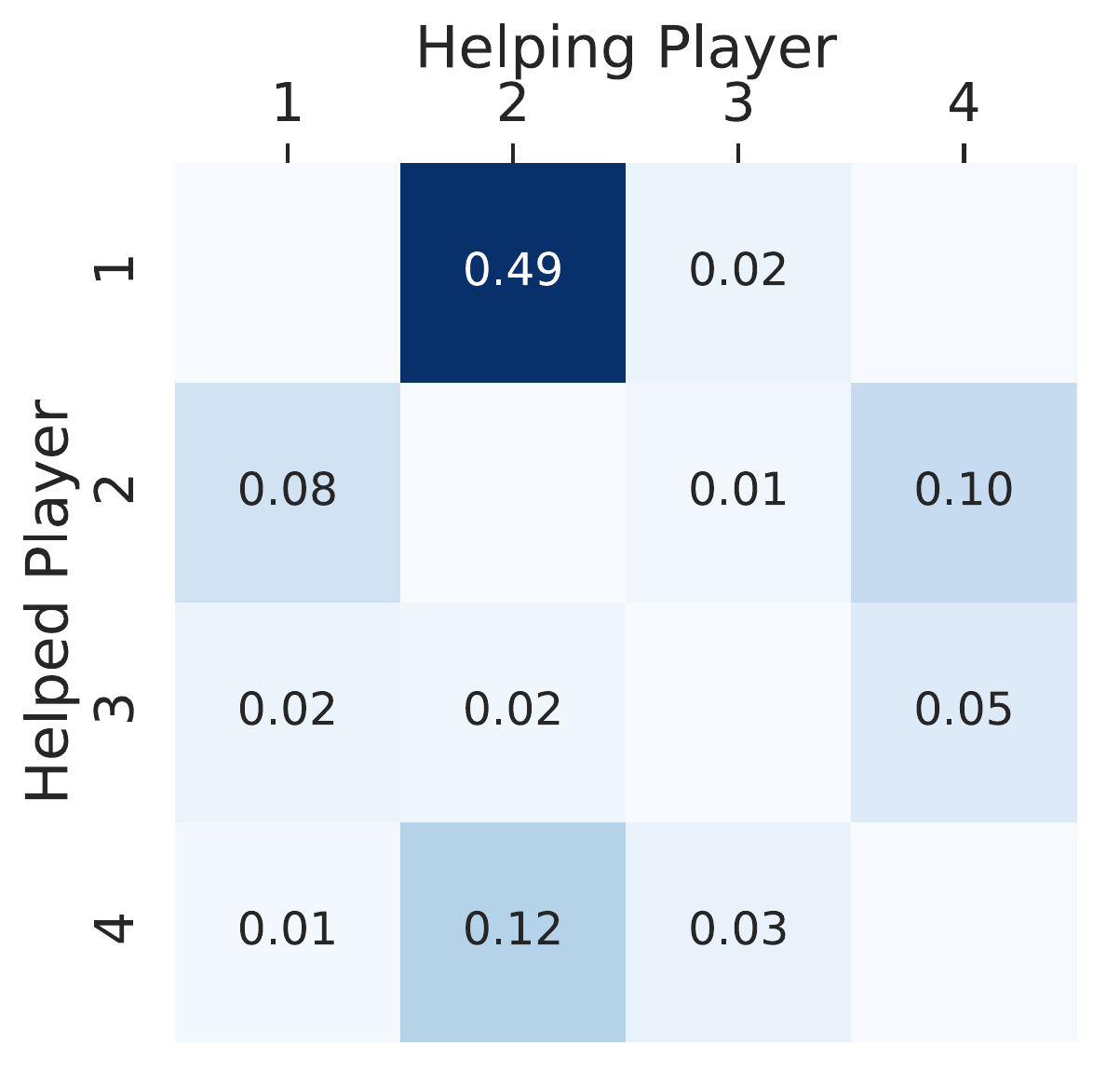}
        \caption{Care matrix run 2.}\label{fig:teqrun2}
    \end{subfigure}
    \;
    \begin{subfigure}{0.23\textwidth}
        \centering
        \includegraphics[width=\linewidth]{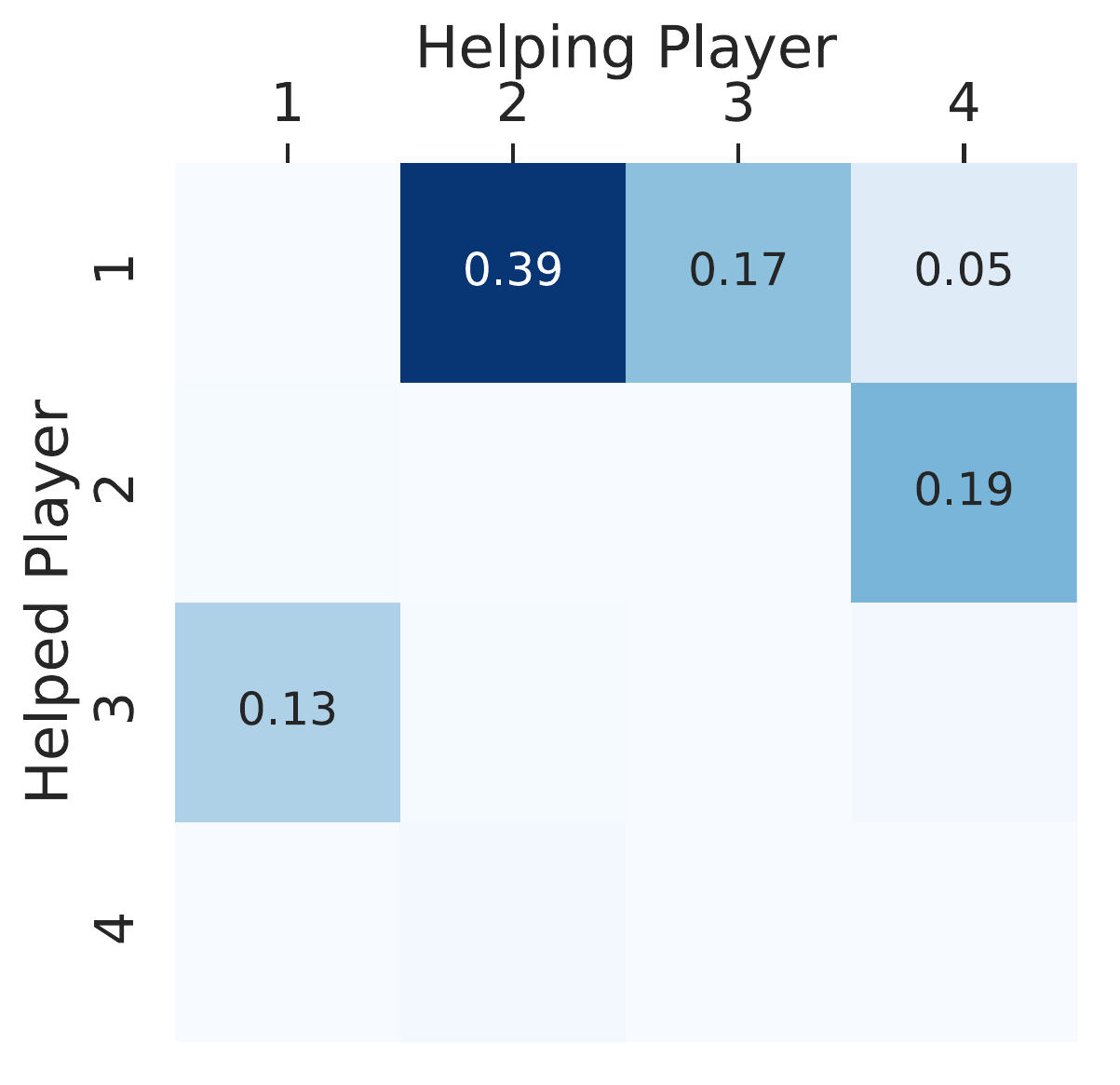}
        \caption{Care matrix run 3.}\label{fig:teqrun3}
    \end{subfigure}
    \;
    \begin{subfigure}{0.23\textwidth}
        \centering
        \includegraphics[width=\linewidth]{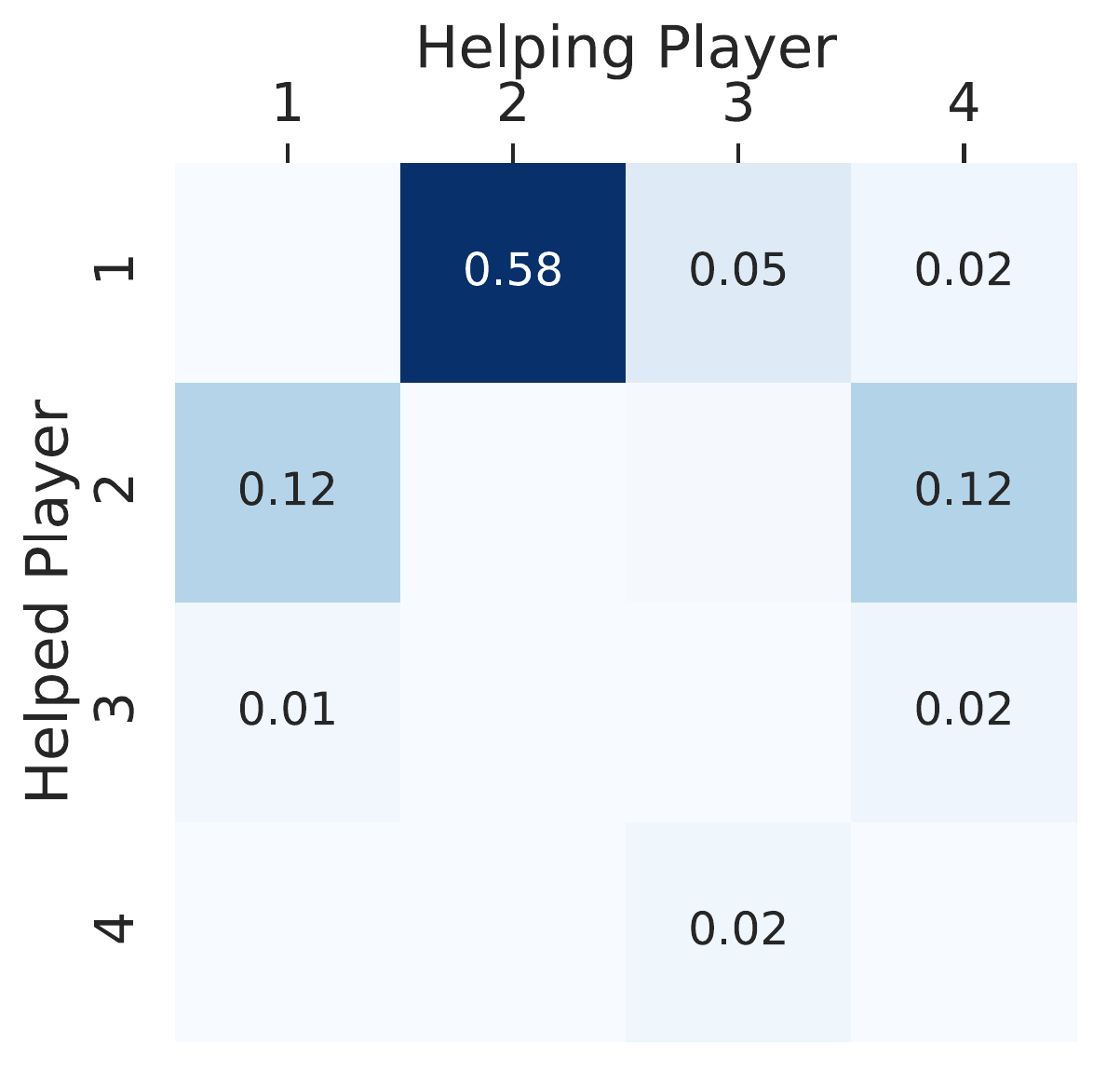}
        \caption{Care matrix run 4}\label{fig:teqrun4}
    \end{subfigure}\\
    \begin{subfigure}{0.23\textwidth}
        \centering
        \includegraphics[width=\linewidth]{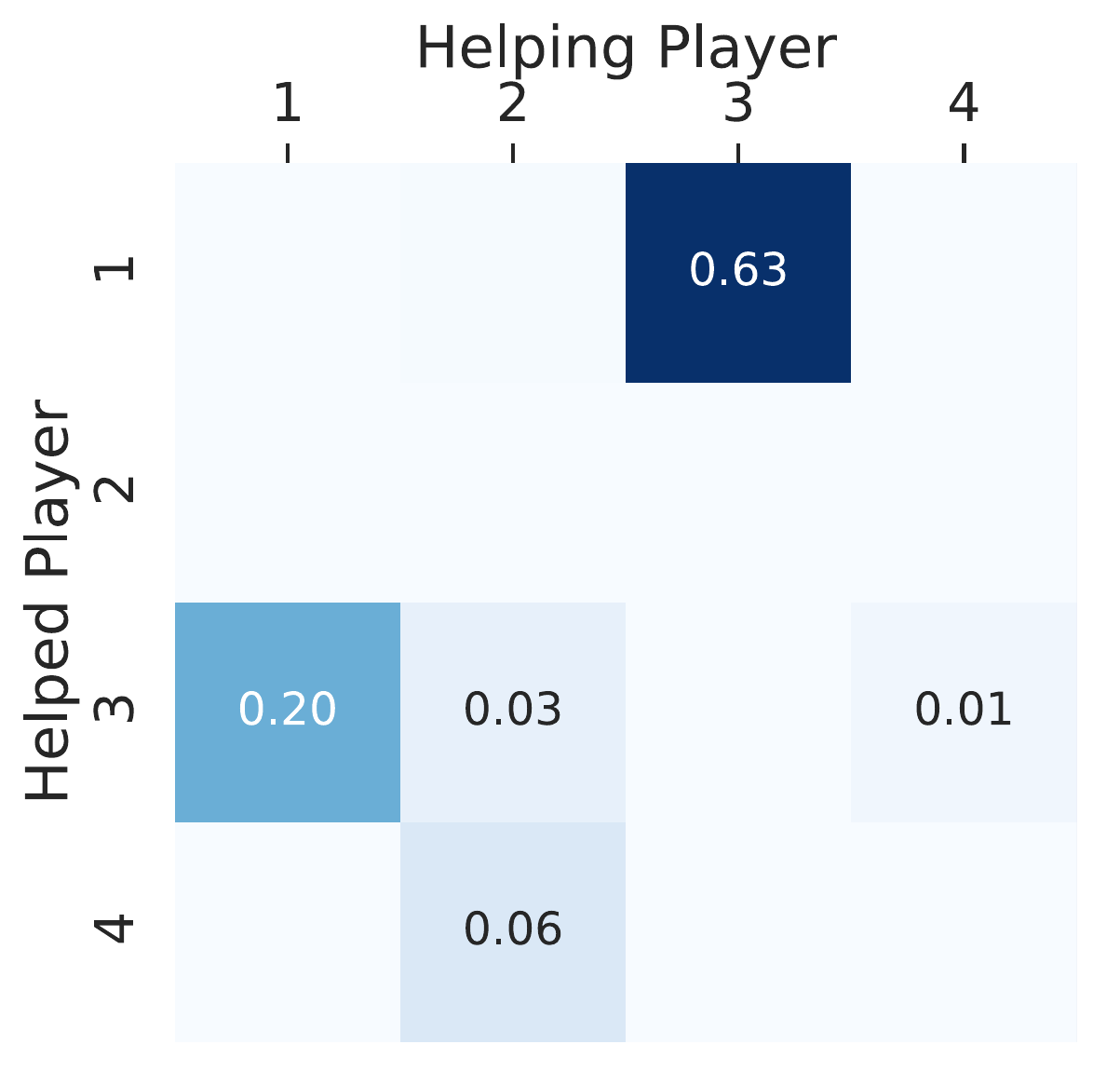}
        \caption{Care matrix run 5.}\label{fig:teqrun5}
    \end{subfigure}
    \;
    \begin{subfigure}{0.23\textwidth}
        \centering
        \includegraphics[width=\linewidth]{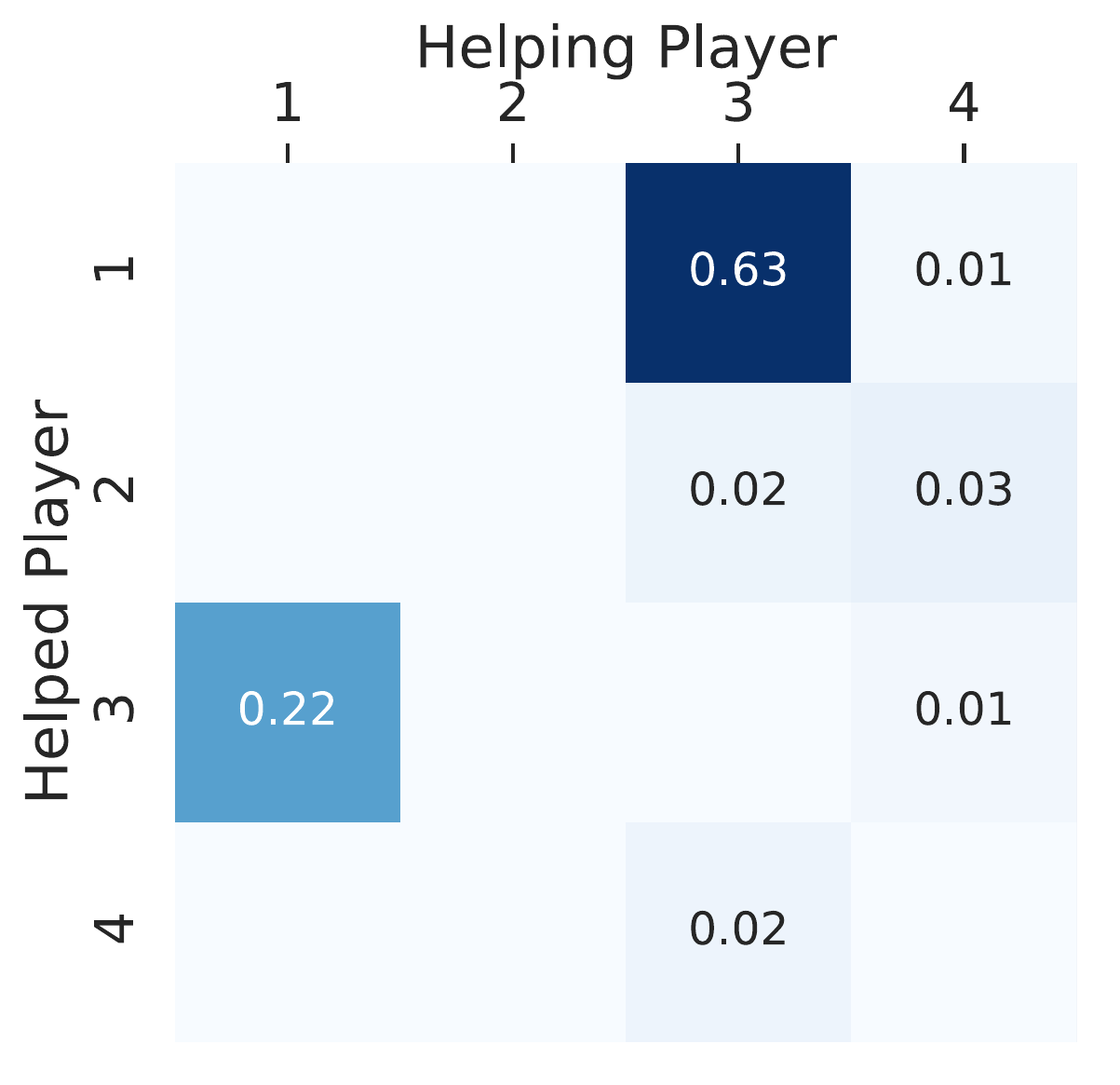}
        \caption{Care matrix run 6.}\label{fig:teqrun6}
    \end{subfigure}
    \;
    \begin{subfigure}{0.23\textwidth}
        \centering
        \includegraphics[width=\linewidth]{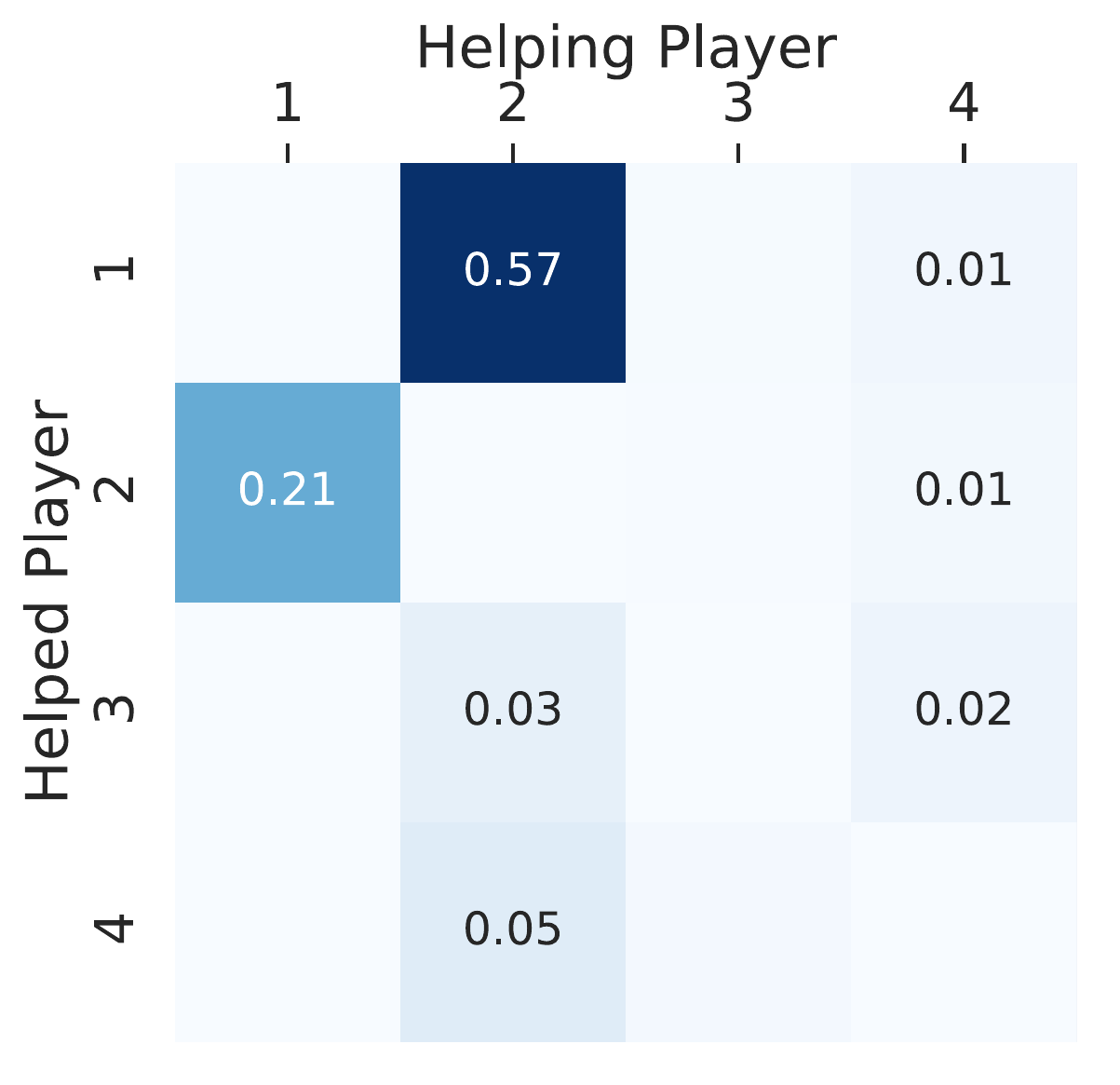}
        \caption{Care matrix run 7.}\label{fig:teqrun7}
    \end{subfigure}
    \;
    \begin{subfigure}{0.23\textwidth}
        \centering
        \includegraphics[width=\linewidth]{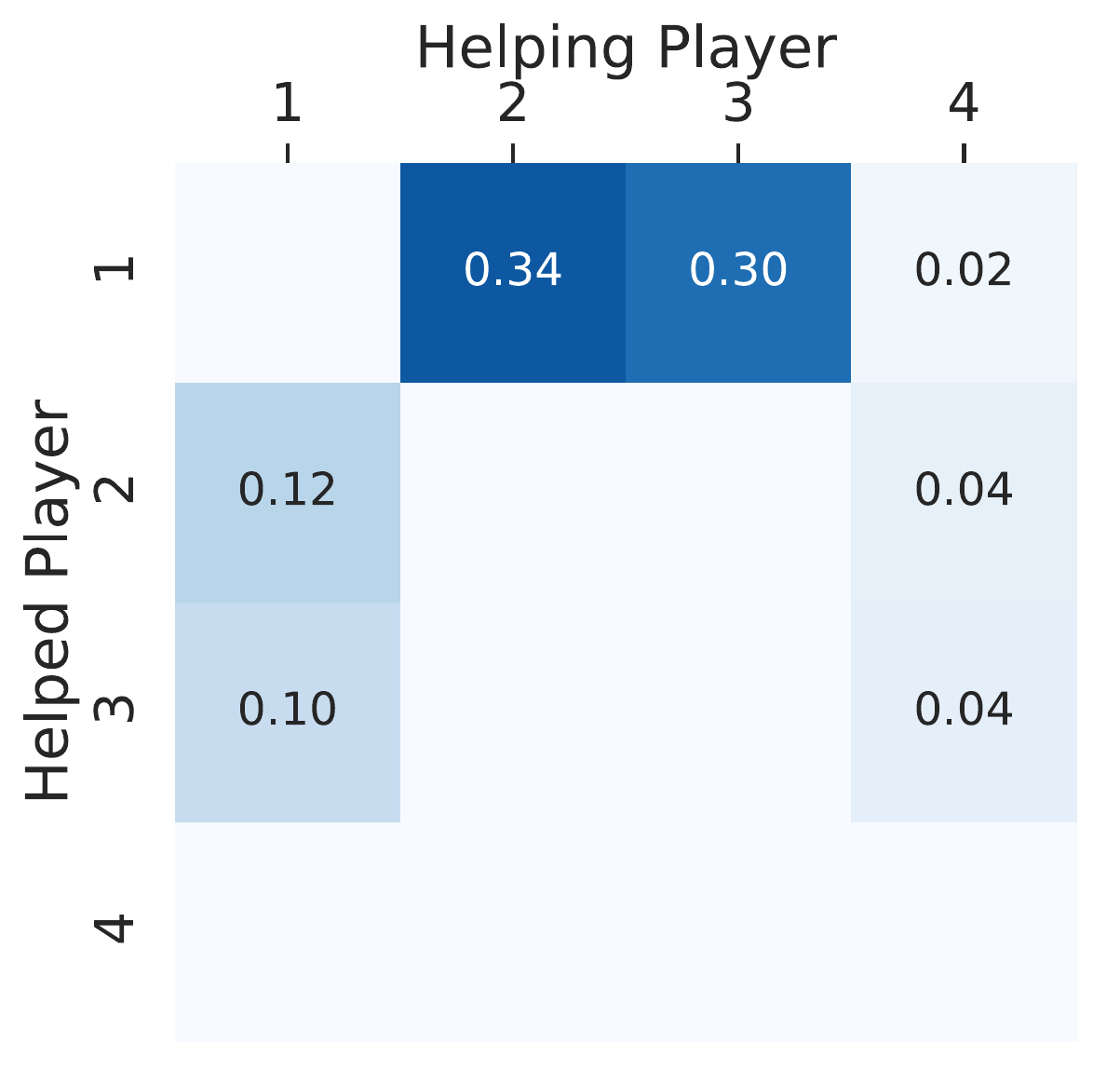}
        \caption{Care matrix run 8.}\label{fig:teqrun8}
    \end{subfigure}
    \caption{Care matrices after convergence for eight individual runs (random seeds) of training in the third t-junction environment (the environment visualized in Figure \ref{fig:t3} in the main paper).}
    \label{fig:care_tjunction}
\end{figure*}

In the third tree-like supply chain that we discussed in Section \ref{sec:topology} in the main paper, we observed a high variance in the individual rewards for agents 2, 3 and 4 (see Table \ref{tab:tjunctions}) because there are multiple stable outcomes that the system randomly converges to. We show these outcomes in terms of care matrices for eight individual runs (random seeds) in Figure \ref{fig:care_tjunction}. Note first that there are indeed large differences between the care matrices across individual runs. Moreover, we see in most care matrices (Figures \ref{fig:teqrun4} to \ref{fig:teqrun8}) that, despite the fact that different agents are involved, similar patterns of reciprocal care emerge as we observed in other experiments. Finally, we observe that only in the last run (Figure \ref{fig:teqrun8}), a pair is formed between the first agent and both agent 2 and 3 while for the other runs pairing happens with only one of these center agents.

\end{document}